\documentclass[english,notitlepage,prl,twocolumn]{revtex4-1}
\usepackage[T1]{fontenc}
\usepackage[final]{changes}
\usepackage{amsmath}
\usepackage{amssymb}
\usepackage{graphicx}
\usepackage{babel}
\usepackage{cancel}
\usepackage[colorlinks,urlcolor=cyan,citecolor=blue,linkcolor=magenta]{hyperref}
\usepackage{xcolor}

\begin{document}

\title{Spin-orbital-angular-momentum-coupled quantum gases}
\author{Shi-Guo Peng, and Kaijun Jiang}
\affiliation{State Key Laboratory of Magnetic Resonance and Atomic and Molecular
Physics, Innovation Academy for Precision Measurement Science and
Technology, Chinese Academy of Sciences, Wuhan 430071, China}
\affiliation{Center for Cold Atom Physics, Chinese Academy of Sciences, Wuhan 430071,
China}
\author{Xiao-Long Chen, and Ke-Ji Chen}
\affiliation{Department of Physics and Key Laboratory of Optical Field Manipulation of Zhejiang Province, Zhejiang Sci-Tech University, Hangzhou 310018, China}
\author{Peng Zou}
\affiliation{College of Physics, Qingdao University, Qingdao 266071, China}
\author{Lianyi He}
\affiliation{Department of Physics and State Key Laboratory of Low-Dimensional
Quantum Physics, Tsinghua University, Beijing 100084, China}
\date{\today}

\begin{abstract}
We briefly review the recent progress of theories and experiments
on spin-orbital-angular-momentum (SOAM)-coupled quantum gases. The
coupling between the intrinsic degree of freedom of particles and
their external orbital motions widely exists in \added{the} universe, and leads
to a broad variety of fundamental phenomena \replaced{in both}{both in} \deleted{the} classical
physics and quantum mechanics. \replaced{The recent}{Recent} realization of synthetic SOAM
coupling in cold atoms has attracted a great deal of attention, and
\replaced{stimulated}{stimulates} a large amount of considerations on exotic quantum phases
in both Bose and Fermi gases. In this review, we present a basic idea
of engineering SOAM coupling in neutral atoms, starting from a semiclassical
description of atom-light interaction. Unique features of \deleted{the} single-particle
physics in the presence of SOAM coupling are discussed. The intriguing
ground-state quantum phases of weakly interacting Bose gases are introduced,
with emphasis on a so-called angular stripe phase, which has \added{not} yet been
observed at present. It is demonstrated how to generate a stable giant
vortex in a SOAM-coupled Fermi superfluid. We also discuss \added{the} topological
characters of a Fermi superfluid in the presence of SOAM coupling. We then
introduce the experimental achievement of SOAM coupling in $^{87}$Rb
Bose gases and its first observation of phase transitions. The most
recent development of SOAM-coupled Bose gases in experiments is also
summarized. Regarding the controllability of ultracold quantum gases,
it opens a new era, \replaced{from}{on} the quantum simulation point of view, to study
the fundamental physics \replaced{resulting}{resulted} from SOAM coupling as well as newly
emergent quantum phases.
\end{abstract}
\maketitle

\tableofcontents
\section{Introduction}
\label{sec:introduction}

The spin-orbital-angular-momentum (SOAM) coupling, the coupling between
the spin degree of freedom and the external orbital motion, is one
of the most common phenomena in our nature. A prominent example in
\deleted{the} classical physics is the astronomical fact that the Moon always
shows the same side to the Earth, known as the tidal locking or the
spin-orbit locking~\cite{Barnes2010F}. In \deleted{the} atomic physics, SOAM
coupling is a relativistic effect, which gives rise to the fine structure
of energy levels of hydrogen atoms~\cite{Landau2007Q}. A similar
effect occurs for protons and neutrons moving inside the nucleus,
leading to a shift in their energy levels in the nucleus shell model
\citep{Talmi1963N}. In \deleted{the} condensed-matter physics, an analogous
coupling between the \replaced{electron}{electron's} spin and its velocity, namely the
spin-linear-momentum (SLM) coupling or spin-orbit (SO) coupling,
 results in a variety  of intriguing and fundamental
phenomena, such as the spin\added{-}Hall effect or topological insulators
\citep{Qi2010T}, which have potential applications in quantum devices.
Although numerous fascinating behaviors of many-body quantum systems
are closely related to SOAM or SO coupling, they are mostly intractable
or manifest themselves under extreme conditions. Therefore, it is
of important significance to find a system, which could mimic the
unique features of SOAM- or SO-coupled quantum systems in a controllable
way.

Owing to the advances \replaced{in}{of} the experimental technique, ultracold atomic
gases acquire a high degree of controllability and tunability in interatomic
interaction, geometry, purity, atomic species, and lattice constant
(of optical lattices)~\cite{Bloch2008M,Giorgini2008T,Kohler2006P,Chin2010F,Gross2017Q}.
To date, ultracold quantum gases have emerged as a versatile platform
for exploring a broad variety of many-body phenomena and can realize
physical effects with analogs throughout physics~\cite{Qi2010T,Dalibard2011A,Abanin2019C}.
However, unlike charged particles, neutral atoms cannot experience
the influence of external electromagnetic fields. Fortunately, thanks
to the controlling of atom-light interaction, \replaced{the internal hyperfine states of neutral atoms, 
playing a role of (pseudo-)spin, are coupled to their linear momentum of center-of-mass motion, 
which equivalently introduces a class of SO couplings experienced by atoms.} {a class of synthetic
gauge fields can be engineered in neutral cold atoms. The internal
hyperfine states of neutral atoms play a role of (pseudo-)spin, and
are strongly coupled to their linear momentum of the center-of-mass
in the presence of atom-light interaction} \added{For example, the theoretical scheme of realizing a one-dimensional (1D) SO coupling with equal Rashba and Dresselhaus strengths was proposed in cold atoms according to a Raman process with a simple $\Lambda$-type configuration~\cite{liu2009E}. Then the basic idea was broadly applied in experiments with both bosonic and fermionic atoms~\cite{Lin2011S,Williams2013S,Wang2012S,Cheuk2012S,Zhang2012C,Olson2014T,Fu2014P,Ji2014E,Ji2015S,Hamner2015S,Jimenez-Garcia2015T,Burdick2016L,Song2016S,Li2016S,Livi2016S}. Soon, an impressive amount of theoretical and experimental efforts have been devoted to the realization of high-dimensional SO coupling~\cite{Osterloh2005C,Ruseckas2005N,Juzeliunas2010G,Campbell2011R,Sau2011C,Anderson2013M,Xu2013A,Liu2014R,Anderson2013S,Lu2020I,Wang2018D,Huang2016E,Wu2016R,Wang2021R}. The high-dimensional SO coupling corresponds to a non-Abelian gauge field and has nontrivial geometric or topological effects, which are absent in systems with 1D SO coupling.} Regarding the controllability of ultracold quantum gases, it opens
a new era, \replaced{from}{on} the quantum simulation point of view, to study the fundamental
physics \replaced{resulting}{resulted} from SO-coupling as well as newly emergent quantum
phases~\cite{Zhai2012S,Zhai2015D,Zhang2018S}.

\replaced{Though the SO coupling}{Despite the realization of the SLM coupling
in cold atoms as it} has \replaced{intensively been}{been intensively} studied in the field of condensed
matter\deleted{s} \added{as well as in ultracold atoms}, it is different from the original meaning of SO coupling
in \deleted{the} atomic physics, in which it indicates the coupling between
the spin and the orbital angular momentum. Recently, such type of
SO coupling, i.e., the SOAM coupling, is theoretically proposed in
cold atoms and enriches our understanding of quantum many-body physics
\citep{Liu2006G,DeMarco2015A,Sun2015S,Qu2015Q,Hu2015H,Chen2016S,Vasic2016E,Hou2017A}.
It is experimentally achieved in $^{87}$Rb Bose gases according to
a Raman process by using a pair of copropagating laser beams operated
in Laguerre-Gaussian (LG) modes~\cite{Chen2018S,Chen2018R,Zhang2019G}.
The ground-state phase diagram of \added{the} systems is confirmed. The hysteresis
loop is observed across the phase boundary, which is a hallmark of
the first-order phase transitions. This is due to the unique property
of the quantized angular momentum, unlike that of the linear momentum
in SLM-coupled systems\deleted{, where the phase transitions are mostly the
second-order kind}~\cite{Li2012Q,Martone2012A,Chen2017Q,Chen2018Q}.
Later on, a considerable amount of attention has been paid \replaced{to}{on} a supersolid-like
phase~\cite{Chen2020A,Chen2020Ground,Duan2020S,Chiu2020V,Bidasyuk2022F},
namely the angular stripe phase, which breaks the U(1) gauge symmetry
to behave like a superfluid and also breaks the angular translational
symmetry (or the rotational symmetry) to manifest spatial order in
the angular density~\cite{Boninsegni2012C}. Nevertheless, this kind
of angular stripe phases \replaced{has}{have} \added{not} yet been observed because of the narrow
window of parameters that is hardly reached in experiments at present
\citep{Chiu2020V}.

\replaced{Soon}{Before long}, the idea of SOAM coupling is theoretically generalized
to fermionic systems~\cite{Chen2020G,Wang2021E}, where the pairing
mechanics plays a crucial role \replaced{in}{on} the Fermi superfluid. While it is
shown that the SOAM coupling leads to the spin-dependent vortex formation
in Bose\added{-Einstein} condensates, SOAM coupling alone does not induce vortices
in a Fermi superfluid, since fermions in a Cooper pair would acquire
opposite orbital angular momenta that cancel each other, yielding
a superfluid devoid of vortices. However, by introducing a \replaced{moderate}{modurate}
detuning away from the two-photon resonance in the Raman process,
which breaks the time-reversal symmetry of systems, a giant vortex
superfluid phase could remain stable in SOAM-coupled Fermi gases.
The Cooper pairs can possess quantized angular momenta, featured as
an angular analog of the Fulde-Ferrell pairing state in SO-coupled
Fermi gases, where Cooper pairs inevitably carry finite center-of-mass
momentum due to the asymmetry of SO-dressed Fermi surface under Zeeman
field~\cite{Dong2013F,Shenoy2013F,Wu2013U,Qu2013T,Zhang2013T,Chen2013I,Liu2013T}.
These SOAM-coupling induced vortices have fascinating and unique features.
For example, their size could be as strikingly large as the length
scale comparable to the waist of Raman beams. This is markedly different
from previously studied vortices in atomic Fermi superfluids, where
changes in the vortex-core structure predominantly take place within
a short\added{-}length scale set by the interatomic separation~\cite{Sensarma2006V,Chien2006G}.
Besides, the vortex core exhibits a large spin imbalance, which originates
from spin-polarized vortex\added{-}bound states, or the so-called Caroli--de
Gennes--Matricon (CdGM) states~\cite{Caroli1964B}, and would serve
as an ideal experimental signal. Subsequently, the topological characters
of a ring-shape SOAM-coupled Fermi superfluid are explored, in which
the ac-stark potential of Raman beams provides a strong radial confinement
\citep{Chen-22}. The genic features of topological superfluid are
encoded in the quantized angular degree of freedom. Nevertheless,
a fundamental hurdle to the experimental observation of the SOAM-coupling\added{-}induced pairing states is the inevitable heating during the Raman
process, which makes it difficult to cool the system below the superfluid \added{transition}
temperature. Instead, it is reasonable to expect that molecule states
in the SOAM-coupled Fermi gases could survive even above the critical
temperature, as what occurs in SO-coupled systems~\cite{Williams2013R,Fu2014P}.
Accordingly, a scenario, \deleted{by} using the radio-frequency spectroscopy
based on \deleted{the} two-body physics, provides an accessible detection of
the pairing mechanics under current experimental conditions~\cite{Han2022M}.

The rest of the review is arranged as follows. In the next section,
we present the theoretical scheme \added{of} how the SOAM-coupling effect could
be achieved in cold atoms according to a Raman transition. Starting
from a semiclassical description of atom-light interaction, the effective
Hamiltonian of a single atom is derived. The possible symmetries of
the system \replaced{are}{is} then discussed, which play an important role in understanding \added{the}
fundamental properties of the system. In Sec.~\ref{sec:SingleParticle}, \deleted{the} single-particle
physics is introduced, including the energy spectrum and the intriguing
spin texture of the ground state. The emergent synthetic gauge field
experienced by cold atoms in the presence of atom-light interaction
is also demonstrated. In Sec.~\ref{sec:SOAMCBose}, the ground-state properties of weakly
interacting SOAM-coupled Bose gases are discussed, based on the solution
of \added{the} Gross-Pitaevskii equation. The quantum phases of Bose condensates
are then identified as well. The emphasis on the properties of the
angular stripe phases is presented. We also propose \deleted{the} possible scenarios
to enlarge the window of parameters that might be accessible in experiments
to observe them. It follows the discussion on the ground state of
interacting SOAM-coupled Fermi gases in Sec.~\ref{sec:SOAMCFermi}. The pairing mechanism of fermions
in the presence of SOAM-coupling is introduced. Two exotic pairing states, i.e., SOAM-coupling-induced
giant vortex and topological superfluid states, are shown to be stabilized under SOAM coupling. Subsequently,
the most recent progress of experiments related to SOAM-coupled quantum gases is summarized in Sec.~\ref{sec:Experiments}.
Sec.~\ref{sec:outlook} is devoted to the conclusions and outlooks for future advances.

\section{Spin-orbital-angular-momentum coupling}
\label{sec:SOAMCHamiltonian}

\begin{figure}
\includegraphics[width=0.6\columnwidth]{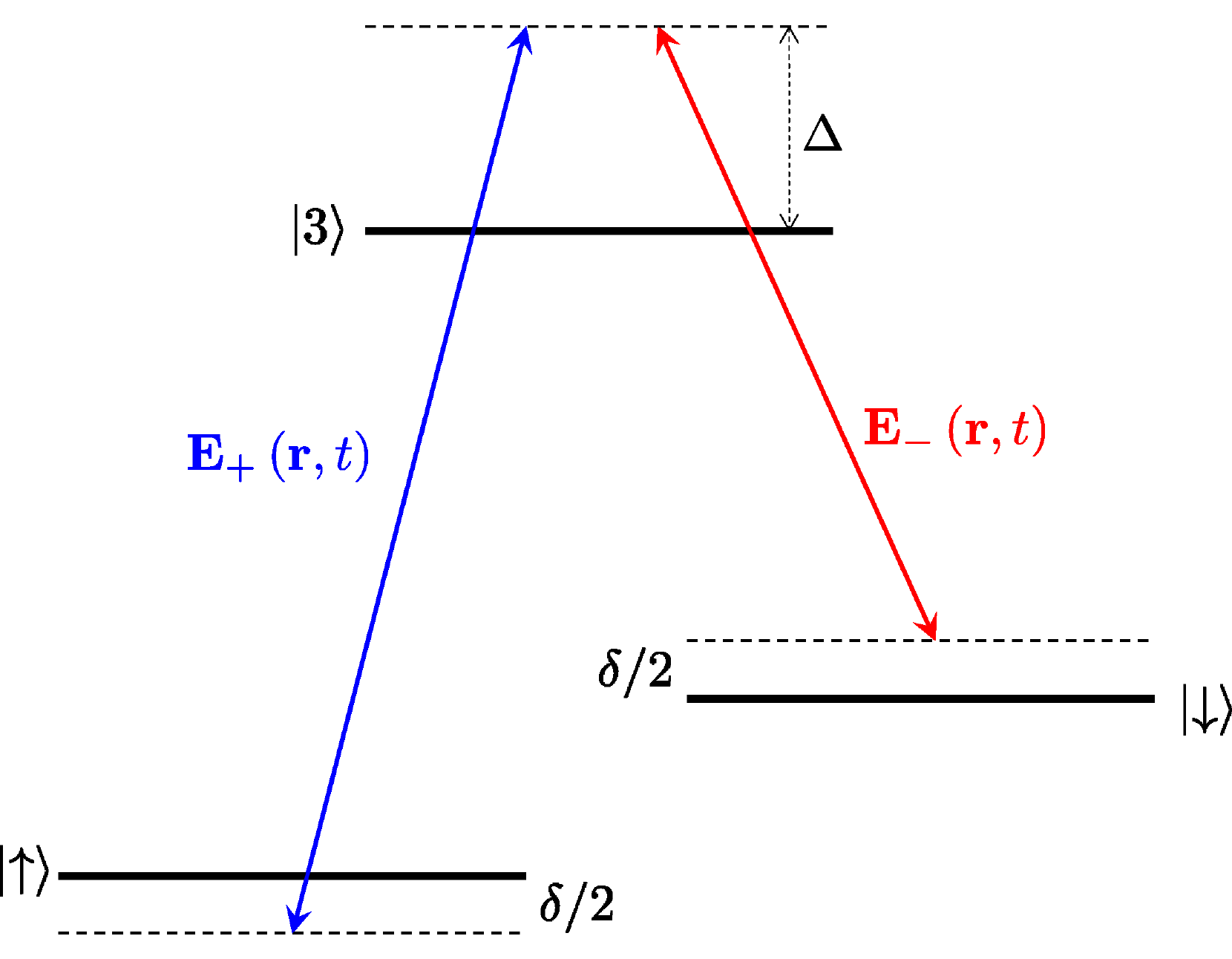}
\caption{Scheme of the SOAM coupling. Two hyperfine states of atoms (denoted
by $\left|\uparrow\right\rangle $ and $\left|\downarrow\right\rangle $)
are coupled to an excited state $\left|3\right\rangle $
by a pair of far-detuned copropagating Laguerre-Gaussian beams ${\bf E}_{\pm}\left({\bf r},t\right)$.
Here, $\delta$ and $\Delta$ are the \replaced{two}{single}-photon and
\replaced{single}{two}-photon detunings respectively.}
\label{fig:Fig1RamanProcess}
\end{figure}

The key idea of generating SOAM coupling in neutral atoms has been
proposed by several earlier works~\cite{DeMarco2015A,Sun2015S,Qu2015Q},
in which two hyperfine states of atoms are coupled by a pair of copropagating
LG beams according to a Raman process as shown in Fig.~\ref{fig:Fig1RamanProcess}.
The two LG beams carry different orbital angular momenta along the
direction of beam propagation, that leads to an orbital angular momentum
change of atoms when transitioning between the two ground hyperfine
states. In the \replaced{following}{follows}, we are going to derive the effective Hamiltonian
of an alkali metal atom in the presence of a pair \added{of} Raman LG beams based on the semiclassical
description of atom-light interaction~\cite{Scully2008Q,Liu2006G,Marzlin1997V}.

\subsection{Semiclassical theory of atom-light interactions}

In the semiclassical theory, the atom-light interaction during the
Raman process is described by the following Hamiltonian
\begin{equation}
\hat{H}_{al}=\left[\begin{array}{ccc}
\hbar\omega_{\uparrow} & 0 & V_{\uparrow3}\\
0 & \hbar\omega_{\downarrow} & V_{\downarrow3}\\
V_{3\uparrow} & V_{3\downarrow} & \hbar\omega_{3}
\end{array}\right]\label{eq:SemiTheory1}
\end{equation}
in the bare hyperfine basis $\left[\left|\uparrow\right\rangle ,\left|\downarrow\right\rangle ,\left|3\right\rangle \right]^{T}$,
where $\hbar\omega_{\sigma}$ ($\sigma=\uparrow,\downarrow,3$) is
the bare energy of atoms in different hyperfine states, and $V_{\sigma\sigma^{\prime}}=-\left\langle \sigma\right|{\bf d}\cdot{\bf E}_{(\pm)}\left|\sigma^{\prime}\right\rangle $
characterizes the dipole interactions between the valence electron
and light fields. Here, ${\bf d}$ is the electric moment of the valence
electron and
\begin{equation}
{\bf E}_{\pm}\left({\bf r},t\right)=\frac{1}{2}\hat{{\bf e}}_{\pm}\left[\mathcal{E}_{\pm}\left({\bf r}\right)e^{-i\omega_{\pm}t}+h.c.\right]
\end{equation}
 are the electric fields experienced by atoms, where $\hat{{\bf e}}_{\pm}$
denote the unit vectors of the polarization direction of light, $\omega_{\pm}$
are the angular frequencies of Raman beams, and $\mathcal{E}_{\pm}\left({\bf r}\right)$
are the spatial complex amplitudes. The wave function of atoms can also
be written in the bare hyperfine basis as
\begin{equation}
\Psi\left(t\right)=\left[\begin{array}{c}
c_{\uparrow}\left(t\right)e^{-i\eta_{\uparrow}t}\\
c_{\downarrow}\left(t\right)e^{-i\eta_{\downarrow}t}\\
c_{3}\left(t\right)e^{-i\eta_{3}t}
\end{array}\right],\label{eq:SemiTheory3}
\end{equation}
and $\eta_{\sigma}$ are arbitrary factors that are chosen for later
convenience. Inserting Eqs.~\eqref{eq:SemiTheory1} and~\eqref{eq:SemiTheory3}
into the Schr\"{o}dinger equation $i\hbar\partial_{t}\Psi=\hat{H}_{al}\Psi$
and by choosing
\begin{eqnarray}
\eta_{\uparrow} & = & \frac{1}{2}\left(\omega_{\uparrow}+\omega_{\downarrow}-\omega_{+}+\omega_{-}\right),\\
\eta_{\downarrow} & = & \frac{1}{2}\left(\omega_{\uparrow}+\omega_{\downarrow}+\omega_{+}-\omega_{-}\right),\\
\eta_{3} & = & \frac{1}{2}\left(\omega_{\uparrow}+\omega_{\downarrow}+\omega_{+}+\omega_{-}\right),
\end{eqnarray}
 we easily obtain
\begin{eqnarray}
i\hbar\frac{dc_{\uparrow}}{dt} & = & +\frac{\delta}{2}c_{\uparrow}+\frac{1}{2}\rho_{+}^{*}\mathcal{E}_{+}^{*}\left({\bf r}\right)c_{3},\label{eq:SemiTheory7}\\
i\hbar\frac{dc_{\downarrow}}{dt} & = & -\frac{\delta}{2}c_{\downarrow}+\frac{1}{2}\rho_{-}^{*}\mathcal{E}_{-}^{*}\left({\bf r}\right)c_{3},\label{eq:SemiTheory8}\\
i\hbar\frac{dc_{3}}{dt} & = & \frac{1}{2}\rho_{+}\mathcal{E}_{+}\left({\bf r}\right)c_{\uparrow}+\frac{1}{2}\rho_{-}\mathcal{E}_{-}\left({\bf r}\right)c_{\downarrow}-\Delta c_{3}\label{eq:SemiTheory9}
\end{eqnarray}
under the rotating-wave approximation (RWA)~\cite{Scully2008Q},
where $\rho_{+}\equiv\left\langle 3\right|-{\bf d}\cdot\hat{{\bf e}}_{+}\left|\uparrow\right\rangle $
and $\rho_{-}\equiv\left\langle 3\right|-{\bf d}\cdot\hat{{\bf e}}_{-}\left|\downarrow\right\rangle $
are the matrix elements of the electric dipole moments, and $\delta/\hbar\equiv\left(\omega_{+}-\omega_{-}\right)-\left(\omega_{\downarrow}-\omega_{\uparrow}\right)$
and $\Delta/\hbar\equiv\left(\omega_{+}+\omega_{-}\right)/2-\left[\omega_{3}-\left(\omega_{\uparrow}+\omega_{\downarrow}\right)/2\right]$
are respectively the two-photon and single-photon detunings. When
the excited state $\left|3\right\rangle $ is far from the resonance,
we may adiabatically eliminate it by setting $\partial_{t}c_{3}\left(t\right)=0$,
and then Eq.~\eqref{eq:SemiTheory9} yields
\begin{equation}
c_{3}\approx\frac{\rho_{+}\mathcal{E}_{+}\left({\bf r}\right)}{2\Delta}c_{\uparrow}+\frac{\rho_{-}\mathcal{E}_{-}\left({\bf r}\right)}{2\Delta}c_{\downarrow}.\label{eq:SemiTheory10}
\end{equation}
After inserting Eq.~\eqref{eq:SemiTheory10} into Eqs.~\eqref{eq:SemiTheory7}
and~\eqref{eq:SemiTheory8}, we arrive at
\begin{equation}
i\hbar\frac{d}{dt}\left[\begin{array}{c}
c_{\uparrow}\\
c_{\downarrow}
\end{array}\right]=\left[\begin{array}{cc}
\delta/2+\chi_{+}\left({\bf r}\right) & \Omega\left({\bf r}\right)\\
\Omega^{*}\left({\bf r}\right) & -\delta/2+\chi_{-}\left({\bf r}\right)
\end{array}\right]\left[\begin{array}{c}
c_{\uparrow}\\
c_{\downarrow}
\end{array}\right],
\end{equation}
where
\begin{eqnarray}
\chi_{\pm}\left({\bf r}\right) & \equiv & \frac{\left|\rho_{\pm}\mathcal{E}_{\pm}\left({\bf r}\right)\right|^{2}}{4\Delta},\\
\Omega\left({\bf r}\right) & \equiv & \frac{\rho_{+}^{*}\rho_{-}\mathcal{E}_{+}^{*}\left({\bf r}\right)\mathcal{E}_{-}\left({\bf r}\right)}{4\Delta}
\end{eqnarray}
 are respectively the diagonal and off-diagonal ac-stark shifts. We
find that the diagonal ac-stark shift $\chi_{\pm}\left({\bf r}\right)$
provides an effective trapping potential for atoms that could be removed
by properly choosing a tune-out wavelength of LG beams in the experiment
\citep{Zhang2019G,Holmgren2012M,Herold2012P}. The off-diagonal ac-stark
shift $\Omega\left({\bf r}\right)$ leads to a space-dependent coupling
between two hyperfine states $\left|\uparrow\right\rangle $ and $\left|\downarrow\right\rangle $,
and then results in \replaced{a}{an} SOAM coupling of atoms as we will see
below.

\subsection{SOAM-coupled Hamiltonian and symmetries}

Without loss of generality, we consider a two-dimensional (2D) configuration
for simplicity \added{in} that atoms are confined in the $z=0$ plane, and the
pair of Raman LG beams copropagate along the $-{\bf \it{z}}$ direction
with the spatial complex amplitude
\begin{equation}
\mathcal{E}_{\pm}\left({\bf r}\right)=\sqrt{2I_{0}}e^{il_{\pm}\varphi}\left(\frac{r}{w}\right)^{\left|l_{\pm}\right|}e^{-r^{2}/w^{2}},
\end{equation}
where $I_{0}$ and $w$ are respectively the intensity and waist of
the beams, and $l_{\pm}$ are their winding numbers. Here, we have
\replaced{adopted}{adapted} a \replaced{polar}{cylindrical} coordinate ${\bf r}=\left(r,\varphi\right)$.
Consequently, the single-atom Hamiltonian in the presence of Raman
LG beams takes an effective form of
\begin{equation}
\hat{H}_{0}=-\frac{\hbar^{2}}{2m}\nabla^{2}+V_{ext}\left(r\right)+\hat{H}_{so}\label{eq:1Ham1}
\end{equation}
with
\begin{equation}
\hat{H}_{so}=\left[\begin{array}{cc}
\delta/2+\chi\left(r\right) & \Omega\left(r\right)e^{-i2n\varphi}\\
\Omega\left(r\right)e^{+i2n\varphi} & -\delta/2+\chi\left(r\right)
\end{array}\right],\label{eq:SOCHam}
\end{equation}
where $V_{ext}\left(r\right)=m\omega^{2}r^{2}/2$ is the external
harmonic potential, $n=\left(l_{+}-l_{-}\right)/2$ is the angular
momentum transferred to atoms, and $\Omega\left(r\right)\equiv\Omega_{R}\left(r/w\right)^{\left|l_{+}\right|+\left|l_{-}\right|}e^{-2r^{2}/w^{2}}$,
with the coupling strength $\Omega_{R}=\left|\rho_{+}^{*}\rho_{-}\right|I_{0}/2\Delta$,
is the radial Rabi frequency. Here, we have assumed $\chi_{+}\left(r\right)=\chi_{-}\left(r\right)\equiv\chi\left(r\right)$
for simplicity, which is only dependent on $\left|{\bf r}\right|=r$.
Besides, we also discard the internal phase difference between the
matrix elements of electric dipole moments $\rho_{+}$ and $\rho_{-}$
that is irrelevant to the problem.

Let us then analyse the possible symmetries of the single-atom Hamiltonian
\eqref{eq:1Ham1}. It is obvious that the rotational symmetry with
respect to the $z$ axis is broken by the dependence of Raman coupling
$\hat{H}_{so}$ on the azimuthal angle $\varphi$. The orbital angular
momentum of the atom is thus no longer conserved, since it follows a
change of orbital angular momentum accompanied by the spin flip in
the Raman process~\cite{DeMarco2015A,Chen2020A,Duan2020S}. By introducing
a unitary transformation $\hat{U}=\exp\left(in\varphi\hat{\sigma}_{z}\right)$,
the single-atom Hamiltonian becomes $\hat{\mathcal{H}}_{0}=\hat{U}\hat{H}_{0}\hat{U}^{\dagger}$,
i.e.,
\begin{multline}
\hat{\mathcal{H}}_{0}=-\frac{\hbar^{2}}{2mr}\frac{\partial}{\partial r}r\frac{\partial}{\partial r}+\frac{\left(\hat{L}_{z}-n\hbar\hat{\sigma}_{z}\right)^{2}}{2mr^{2}}\\
+V_{ext}\left(r\right)+\chi\left(r\right)+\Omega\left(r\right)\hat{\sigma}_{x}+\frac{\delta}{2}\hat{\sigma}_{z},\label{eq:1Ham2}
\end{multline}
where $\hat{L}_{z}=-i\hbar\partial/\partial\varphi$, and $\hat{\sigma}_{x,z}$
are Pauli matrices. We find that the angular momentum $\hat{L}_{z}$
commutes with the single-atom Hamiltonian $\hat{\mathcal{H}}_{0}$
under the unitary transformation $\hat{U}$, and thus $\hat{L}_{z}$
is conserved. Therefore, the original single-atom Hamiltonian $\hat{H}_{0}$
demonstrates the symmetry under the redefined rotational transformation
$\hat{\mathcal{R}}\left(\varphi\right)\equiv\hat{U}^{\dagger}e^{-i\hat{L}_{z}\varphi/\hbar}\hat{U}$.
Then $\hat{L}_{z}$ may be regarded as the operator of quasi\added{-}angular
momentum (QAM), and each eigenstate of the system possesses definite
values of QAM characterized by the corresponding quantum number $l_{z}$.
It is \deleted{apparently} related to the angular momentum of each spin component
in the laboratory frame by $l_{\uparrow,\downarrow}=l_{z}\mp n$.
Here, we find that the term of $\hat{L}_{z}\hat{\sigma}_{z}$ appears,
which characterizes the crucial SOAM-coupling effect that is well
familiar to us in \deleted{the} atomic physics.

At the resonance of Raman coupling with $\delta=0$, the system demonstrates
an additional symmetry, namely the time-reversal symmetry. The single-atom
Hamiltonian $\hat{H}_{0}$ commutes with the time-reversal operator
$\hat{T}=\hat{\sigma}_{x}\hat{K}$, and thus is invariant under the
time-reversal transformation, where $\hat{K}$ denotes the operator
of complex conjugation~\cite{DeMarco2015A,Duan2020S}. This symmetry
may be translated into \added{a} QAM frame as $\hat{\mathcal{T}}=\hat{U}\hat{T}\hat{U}^{\dagger}$,
and then we have $\left[\hat{\mathcal{T}},\hat{\mathcal{H}}_{0}\right]=0$.
The time-reversal symmetry guarantees that the spectrum of a single
atom is symmetric about the QAM $l_{z}=n$.

\section{Single-particle physics}
\label{sec:SingleParticle}

\subsection{Single-particle spectrum}\label{sec:SPS}

\begin{figure}
\includegraphics[width=1\columnwidth]{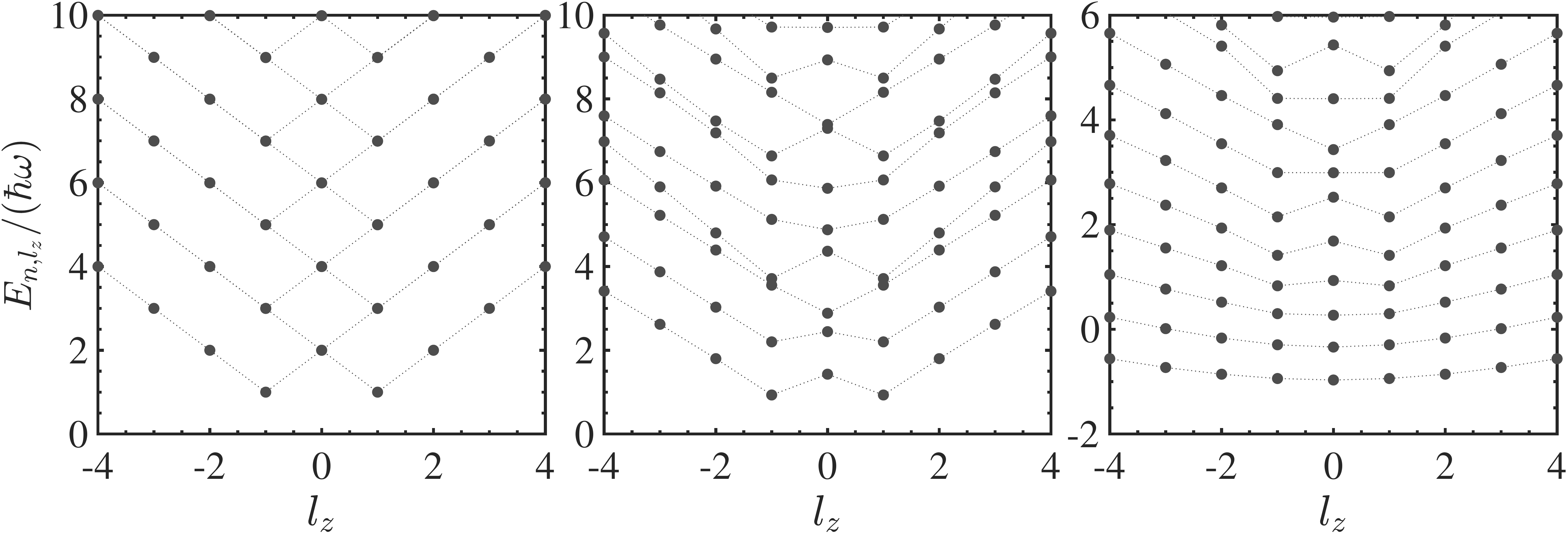}
\caption{The single-particle dispersion for three typical SOAM-coupling strength
$\Omega_{R}/\hbar\omega=0$, $100$ and $250$ (left to right). The
energy is characterized by two quantum numbers, i.e., the radial quantum
number or band index $n$ and the quasi angular momentum $l_{z}$.
Here, we have set the two-photon detuning $\delta=0$ and the off-diagonal
ac-stark shift $\chi\left(r\right)$=0 as well.}
\label{fig:Fig2SingleAtomSpectrum}
\end{figure}

Regarding the rotational symmetry of the single-atom Hamiltonian \eqref{eq:1Ham2},
each eigenstate of a single atom should possess a definite QAM $l_{z}$,
whose wave function may be written as
\begin{equation}
\Psi_{l_{z}}\left({\bf r}\right)=\left[\begin{array}{c}
\psi_{\uparrow}\left(r\right)\\
\psi_{\downarrow}\left(r\right)
\end{array}\right]\frac{e^{il_{z}\varphi}}{\sqrt{2\pi}}.
\end{equation}
Then the Schr\"{o}dinger equation $\hat{\mathcal{H}}_{0}\Psi_{l_{z}}\left({\bf r}\right)=E\Psi_{l_{z}}\left({\bf r}\right)$
reduces to two coupled one-dimensional radial equations of $\psi_{\uparrow,\downarrow}\left(r\right)$,
which are easily solved by using \added{the} finite-difference method~\cite{DeMarco2015A}.
The single-particle dispersion relations between the energy $E$ and
QAM $l_{z}$ are shown in Fig.~\ref{fig:Fig2SingleAtomSpectrum} for
three typical coupling strengths. Here, the winding numbers of two
Raman beams are respectively chosen to be $l_{+}=-2$ and $l_{-}=0$,
and we consider the situation at the two-photon resonance with $\delta=0$.
The symmetry of the single-particle dispersion about $l_{z}=0$ is
guaranteed by the time-reversal symmetry $\hat{\mathcal{T}}$ discussed
above. This means that the Schr\"{o}dinger equation is invariant
under the exchange of the spin indices, which sends $n\rightarrow-n$
as well.

In the absence of SOAM coupling, i.e., $\Omega_{R}=0$, the energy
band structure is simply that of the spinor harmonic oscillator with
the excitation interval $\hbar\omega$. The ground state is characterized
by the QAM $l_{z}=\pm1$, which is doubly degenerate. The angular momentum
for each spin component in the laboratory frame is $\left(l_{\uparrow},l_{\downarrow}\right)=\left(0,-2\right)$
and $\left(l_{\uparrow},l_{\downarrow}\right)=\left(2,0\right)$ corresponding
to the QAM $l_{z}=-1$ and $1$, respectively. As the SOAM-coupling strength
$\Omega_{R}$ gradually increases, the spin is no longer a good quantum
number, and small amounts of atoms are transferred into the previously
vacant spin component, while the ground state still has the two-fold
degeneracy with definite QAM. By further increasing the coupling strength,
the system finally jumps into the QAM $l_{z}=0$ ground state, which
gives rise to a first-order phase transition to a spin-balanced population
\citep{Zhang2019G}. This is due to the unique property of the quantized
angular momentum in SOAM-coupled systems. In SLM-coupled systems,
such phase transition is a continuous type, where the doubly degenerate
ground states finally merge into each other as the Raman-coupling
strength increases~\cite{Li2012Q,Martone2012A,Chen2017Q,Chen2018Q}.

\begin{figure}
\includegraphics[width=1\columnwidth]{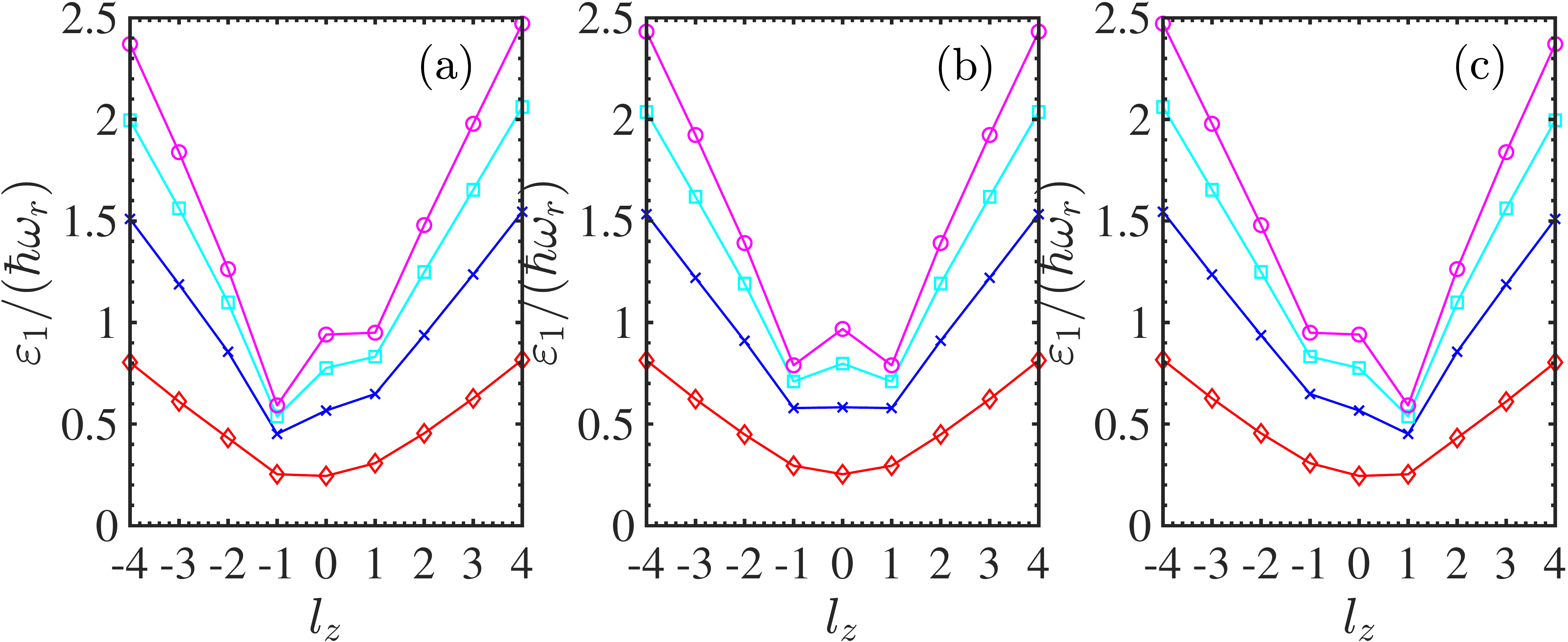}
\caption{Lowest energy band in the single-particle dispersion at (a) negative,
(b) zero, and (c) positive two-photon detuning $\delta$ when the
SOAM-coupling strength \added{$\Omega_{R}$} increases (top to bottom). Adapted from Ref.~\citep{Chen2020A}.}
\label{fig:Fig3LowestBand-SB}
\end{figure}

Let us look closely at the evolution of the lowest-energy band as
presented in Fig.~\ref{fig:Fig3LowestBand-SB} at different coupling
strength\added{s}. Its symmetry about $l_{z}=0$ is preserved by the time-reversal
symmetry at the two-photon resonance $\delta=0$ as shown in Fig.~\ref{fig:Fig3LowestBand-SB}(b).
We demonstrate how the ground state evolves from the doubly degenerate
states ($l_{z}=\pm1$) into a single one ($l_{z}=0$) as the coupling
strength increases. However, away from the two-photon resonance with
$\delta\neq0$, we find that the degeneracy of the ground states at
weak coupling strength is lifted, since the time-reversal symmetry
is broken by the two-photon detuning (as shown in Fig.~\ref{fig:Fig3LowestBand-SB}(a)
and (c)). The ground state is located either at $l_{z}=-1$ or $1$
determined by the sign of the two-photon detuning. This gives rise
to an additional first-order phase transition by continuously varying
the two-photon detuning~\cite{Zhang2019G}. At strong coupling strength,
the system again jumps into the QAM $l_{z}=0$ ground state as we
discussed above.

\subsection{Spin textures}

\begin{figure}
\includegraphics[width=1\columnwidth]{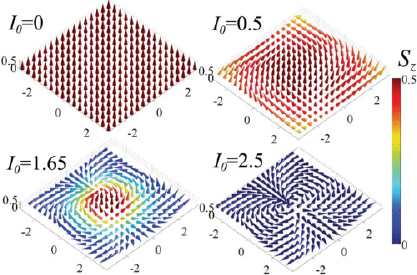}
\caption{Spin texture of the ground state of a SOAM-coupled system at different
coupling strengths. The arrows point in the direction of the local
spin ${\bf S}$, and the color represents the projection of the spin
onto the $z$ axis. The figure is adapted from Ref.~\citep{DeMarco2015A},
as well as corresponding parameters therein.}
\label{fig:Fig4SpinTexture}
\end{figure}

The SO coupling gives rise to intriguing spin textures.
For example, it has been found in studies of 2D Rashba SO-coupled
BECs~\cite{Hu2012S,Ramachandhran2012H} and BECs exposed to LG beams
\citep{Leslie2009C} that the spin texture contains a topological
knot known as a 2D skyrmion. It is one kind of topological defects
protected by their topological nontriviality. Regarding the spin texture
of SOAM-coupled systems, it is useful to define a spin vector~\cite{DeMarco2015A,Hu2012S,Ramachandhran2012H}
\begin{equation}
{\bf S}\left({\bf r}\right)=\frac{\Psi^{\dagger}\left({\bf r}\right)\hat{{\bf s}}\Psi\left({\bf r}\right)}{\left|\Psi\left({\bf r}\right)\right|^{2}},
\end{equation}
with the spin operator $\hat{{\bf s}}=\left(\hbar/2\right)\left(\hat{\sigma}_{x},\hat{\sigma}_{y},\hat{\sigma}_{z}\right)$.
The skyrmion number defined as
\begin{equation}
n_{\text{skyrmion}}=\frac{1}{4\pi}\int{\bf S}\cdot\left(\partial_{x}{\bf S}\times\partial_{y}{\bf S}\right)d{\bf r}
\end{equation}
is a measure of the winding of the spin texture, which distinguishes
a skyrmion texture from that of the vacuum. If it equals \deleted{to} $1$ or
$-1$, a topological knot exists in the spin texture~\cite{Binz2008C,Muhlbauer2009S}.
The ground-state spin texture at four typical coupling strengths is
presented in Fig.~\ref{fig:Fig4SpinTexture}. At weak coupling strength,
the two-fold degenerate ground state gives rise to a spin texture
corresponding to a half skyrmion. As the coupling strength increases,
the ground state finally jumps to the QAM $l_{z}=0$ state. The system
reaches a spin-balanced state, which does not support a skyrmion texture.
The local spin becomes planar and lies in the $x-y$ plane, and thus
the skyrmion number vanishes.

\subsection{Artificial gauge field}

In the \replaced{following}{follows}, we are going to demonstrate how the artificial (synthetic)
gauge field may emerge for neutral atoms in the presence of atom-light
interactions. It is one of \added{the} essential ingredients for the simulation
of charged particles moving in the electromagnetic field by using
cold atoms~\cite{Dum1996G,Visser1998G,Lin2009S,Lin2009B,Dalibard2011A}, \added{which gives rise to a series of intriguing phenomena, such as the spin Hall effect~\cite{Zhu2006S,Liu2007O, Beeler2013T}}. To this end, we may simply rewrite the atom-light Hamiltonian \eqref{eq:SOCHam}
in an explicit form of
\begin{equation}
\hat{H}_{so}={\bf Z}\left({\bf r}\right)\cdot{\bf s}
\end{equation}
with the effective Zeeman field
\begin{equation}
{\bf Z}\left({\bf r}\right)=\frac{2}{\hbar}\left(\Omega\left(r\right)\cos2n\varphi,\Omega\left(r\right)\sin2n\varphi,\frac{\delta}{2}\right).
\end{equation}
We find easily that the Raman-induced atom-light interaction effectively
provides a spin-magnetic interaction equivalent to that for a spin-half
charged particle in a space-dependent Zeeman field ${\bf Z}\left({\bf r}\right)$.
The diagonalization of $\hat{H}_{so}$ simply gives the dressed spin
states \cite{Dalibard2011A}, i.e.,
\begin{equation}
\left|\zeta_{+}\right\rangle =\left[\begin{array}{c}
\cos\left(\theta/2\right)\\
e^{i\phi}\sin\left(\theta/2\right)
\end{array}\right],\;\left|\zeta_{-}\right\rangle =\left[\begin{array}{c}
-e^{-i\phi}\sin\left(\theta/2\right)\\
\cos\left(\theta/2\right)
\end{array}\right]
\end{equation}
with eigenvalues $\pm\hbar\left|{\bf Z}\left({\bf r}\right)\right|/2$,
respectively, and $\tan\phi=\tan2n\varphi$ and $\tan\theta=2\Omega\left(r\right)/\delta$,
which determine the orientation of the effective Zeeman field. It
is obvious that $\left|\zeta_{\pm}\right\rangle $ denotes the states
that the pseudo-spin of the atom aligns along or inversely to the
direction of \added{the} local Zeeman field. In the \replaced{following}{follows}, let us consider the
atom moves slowly  in the space-dependent Zeeman field ${\bf Z}\left({\bf r}\right)$,
and its pseudo-spin follows adiabatically one of the eigenstates of
$\hat{H}_{so}$, namely for example $\left|\zeta_{+}\right\rangle $.
Then the full wave function of the atom can be written as $\left|\Psi\left({\bf r},t\right)\right\rangle =\psi_{+}\left({\bf r},t\right)\left|\zeta_{+}\right\rangle $,
the evolution of which is governed by the Schr\"{o}dinger equation
$i\hbar\partial_{t}\left|\Psi\left({\bf r},t\right)\right\rangle =\hat{H}_{0}\left|\Psi\left({\bf r},t\right)\right\rangle $
under the Hamiltonian \eqref{eq:1Ham1}. Here, $\psi_{+}\left({\bf r},t\right)$
is the spatial wave function characterizing the center-of-mass motion
of the atom in the internal dressed state $\left|\zeta_{+}\right\rangle $.
By projecting the Schr\"{o}dinger equation onto the internal dressed
state $\left|\zeta_{+}\right\rangle $, we obtain easily
\begin{equation}
i\hbar\frac{\partial\psi_{+}}{\partial t}=\left[\frac{\left(\hat{{\bf P}}-{\bf A}\right)^{2}}{2m}+V_{ext}+W+\frac{\hbar\left|{\bf Z}\right|}{2}\right]\psi_{+}.\label{eq:GaugeFieldEquation}
\end{equation}
Two geometric potentials ${\bf A}$ and $W$ emerge in addition to
the external trapping potential $V_{ext}$ and Zeeman energy $\hbar\left|{\bf Z}\right|/2$,
which are respectively the vector potential~\cite{Dalibard2011A}
\begin{equation}
{\bf A}\left({\bf r}\right)\equiv\left\langle \zeta_{+}\right|i\hbar\nabla\left|\zeta_{+}\right\rangle =\frac{\hbar}{2}\left(\cos\theta-1\right)\nabla\phi,
\end{equation}
and the scalar potential
\begin{equation}
W\left({\bf r}\right)\equiv\frac{\hbar^{2}}{2m}\left|\left\langle \zeta_{-}\right|\nabla\left|\zeta_{+}\right\rangle \right|^{2}=\frac{\hbar^{2}}{8m}\left[\left(\nabla\theta\right)^{2}+\sin^{2}\theta\left(\nabla\phi\right)^{2}\right].
\end{equation}
It is obvious that Eq.~\eqref{eq:GaugeFieldEquation} takes \deleted{exactly}
the same form \replaced{as}{of} that for a spin-half particle with \added{a} unit charge ($q=1$)
\citep{Landau2007Q}. The effective magnetic field ${\bf B}\left({\bf r}\right)$
associated with ${\bf A}\left({\bf r}\right)$ is

\begin{equation}
{\bf B}\left({\bf r}\right)=\nabla\times{\bf A}\left({\bf r}\right)=\frac{\hbar}{2}\nabla\left(\cos\theta\right)\times\nabla\phi.
\end{equation}
Here, we should pay special attention to the difference between the
Zeeman field ${\bf Z}\left({\bf r}\right)$ and the effective magnetic
field ${\bf B}\left({\bf r}\right)$: the Zeeman field ${\bf Z}\left({\bf r}\right)$
only lifts the degeneracy of the bare spin states and does not affect
the center-of-mass motion of the atom, while ${\bf B}\left({\bf r}\right)$
is the magnetic field experienced by the center-of-mass motion of
the atom when it moves in the presence of the atom-light interaction
and keeps staying in the internal dressed state $\left|\zeta_{+}\right\rangle $.
In the presence of Raman LG beams, the effective magnetic field ${\bf B}\left({\bf r}\right)$
takes the explicit form of
\begin{equation}
{\bf B}\left({\bf r}\right)=\frac{n\hbar}{r}\left[\frac{\partial}{\partial r}\frac{\delta/2}{\sqrt{\Omega^{2}\left(r\right)+\delta^{2}/4}}\right]\hat{{\bf e}}_{z},
\end{equation}
which is perpendicular to the $x-y$ plane and along the $z$ axis.

\section{SOAM-coupled Bose gases}
\label{sec:SOAMCBose}
The interatomic interaction plays a crucial role in interacting many-body quantum systems.
In this section, we are going to introduce the ground-state quantum phases of weakly-interacting
Bose-Einstein condensates (BECs) in the presence of SOAM coupling, within the framework of the mean-field theory based on
the Gross-Pitaevskii (GP) equation. The interaction gives rise to the appearance of a variety of intriguing
angular stripe phases, which have \added{not} yet been observed in current experiments. The possible theoretical scheme
is proposed for observing angular stripe phases in the $^{41}$K atomic gas with tunable interatomic interactions.

The mean-field Hamiltonian of a weakly interacting Bose gas in the presence of SOAM coupling takes the form of
\begin{multline}\label{eq:total-E}
\hat{H}=\int d{\bf r}\left\{ \left[\begin{array}{cc}
\Psi_{\uparrow}^{*}, & \Psi_{\downarrow}^{*}\end{array}\right]\hat{\mathcal{H}}_{0}\left[\begin{array}{c}
\Psi_{\uparrow}\\
\Psi_{\downarrow}
\end{array}\right]\right.\\
\left.+\frac{g_{\uparrow\uparrow}}{2}\left|\Psi_{\uparrow}\right|^{\color{red}4}+\frac{g_{\downarrow\downarrow}}{2}\left|\Psi_{\downarrow}\right|^{\color{red}4}+g_{\uparrow\downarrow}\left|\Psi_{\uparrow}\right|^{2}\left|\Psi_{\downarrow}\right|^{2}\right\} ,
\end{multline}
where $\hat{\mathcal{H}}_{0}$ is the single-particle Hamiltonian \eqref{eq:1Ham2}, $\Psi\left({\bf{r}}\right)\equiv\begin{bmatrix}\Psi_{\uparrow}\left({\bf{r}}\right),\Psi_{\downarrow}\left({\bf{r}}\right)\end{bmatrix}^T$ is the spinor wave function for the condensate, and $g_{\sigma\sigma'}$ is the effective 2D intra- ($\sigma=\sigma'$) and inter-species ($\sigma\neq\sigma'$) interaction strength.

\subsection{Gross-Pitaevskii theory and a variational approach}

When the interaction is taken into account, the nonlinearity may spontaneously
break\deleted{s} the rotational symmetry of the system. We can no longer assume that the condensate
possesses the definite QAM $l_{z}$. The ground-state solution of \added{the}
GP equation is first attempted by using a split-step imaginary time
evolution \cite{DeMarco2015A}. Three kinds of quantum phases are
identified: two of them still preserve the rotational symmetry as
the many-body analogs of the single-particle ground state, while the
third one is a newly emergent angular stripe phase, which does not
possess a definite QAM and breaks the rotational symmetry.

\begin{figure}[t]
\centering
\includegraphics[width=0.43\textwidth]{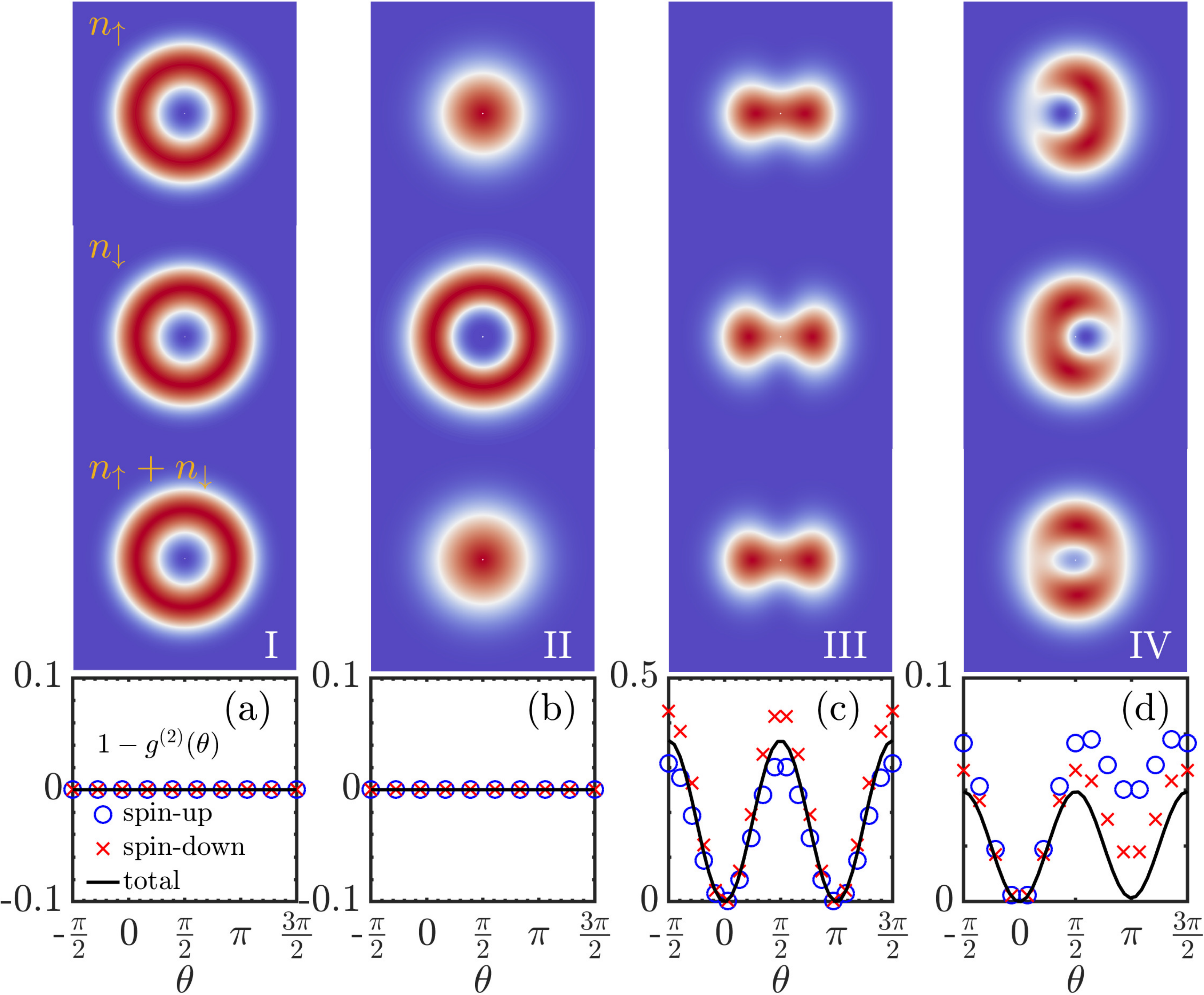}
\caption{(Top three panels) Typical spin-component, total density profiles in four distinct phases. (Bottom panel) In (a-d), the corresponding angular density-density correlation function $1-g^{(2)}(\theta)$ is shown. The results for spin-up, spin-down and total density are indicated by circles, crosses and solid lines, respectively. \added{Here, I is for the vortex-antivortex pair phase, II is for the half-skyrmion phase, III and IV are for angular stripe phases.} Adapted from Ref.~\cite{Chen2020A}.}
\label{fig_n_g2}
\end{figure}

Therefore, apart from the angular stripe phase, one may take the states with a definite QAM for the condensates as a good starting point when taking into account interactions. If the system or the condensate wave function preserves the axial symmetry, one may consider the QAM $l_{z}$ as a good quantum number and adopt the {\it Ansatz} for the condensate~\cite{Chen2020A}

\begin{equation} \label{eq:definite-am ansatz}
    \psi_{_{l_{z}}}\left({\bf{r}}\right)=\begin{bmatrix} f_{_{l_{z}\uparrow}}(r)\\ f_{_{l_{z}\downarrow}}(r)  \end{bmatrix}\frac{e^{il_{z}\varphi}}{\sqrt{2\pi}},
\end{equation}
with the radial wave function $f_{_{l_{z}\sigma}}$, and substitute it into the GP equation,
\begin{equation}  \label{eq:GP-eq-SOAMC}
\begin{bmatrix}
\hat{\mathcal{H}}_{0}(\hat{L}_z)+\mathrm{diag}(\mathcal{L}_\uparrow,\mathcal{L}_\downarrow)
\end{bmatrix} \psi_{_{l_{z}}}= \mu \psi_{_{l_{z}}},
\end{equation}
where $\mu$ is the chemical potential, and the diagonal element takes the form of $\mathcal{L}_{\sigma}\equiv g_{\sigma\sigma}n_{\sigma}+g_{_{\uparrow\downarrow}}n_{\bar{\sigma}}$ (spin index $\sigma\neq\bar{\sigma}$). Thus, one can then determine the ground-state wave function $\psi_{_{l_{z}}}$ of the rotationally symmetric phases at zero temperature by solving self-consistently this GP equation.

Nonetheless, to capture the exotic angular stripe phase that does not possess a definite QAM, a variational \textit{Ansatz} with various angular momenta is adopted for the condensate wave function in determining the ground state with \replaced{an}{a} energy-minimizing method~\cite{Chen2020A}.

In the realistic configuration of recent experiments~\cite{Chen2018S,Zhang2019G}, the orbital angular momentum transferred by the LG lasers is $n=\pm1$, giving rise to minima with the QAM located at $l_z=\pm1$, or $0$ in the single-particle dispersion. Thus, the adopted variational \textit{Ansatz} could be taken as
\begin{equation}  \label{eq:2D-variationalansatz}
    \Psi\left({\bf r}\right)=\alpha e^{i\theta_\alpha}\psi_{_{-1}}+\beta e^{i\theta_\beta}\psi_{_{0}}+\gamma e^{i\theta_\gamma}\psi_{_{1}},
\end{equation}
with real weighting coefficients $\alpha, \beta, \gamma$ and the associated phases $\theta_{\alpha}$, $\theta_{\beta}$, and $\theta_{\gamma}$. In general, $\psi_{_{l_{z}}}$ is the state with a definite QAM $l_{z}$, which can either be the single-particle state or the one obtained by self-consistently solving the GP equation~\eqref{eq:GP-eq-SOAMC} when the interaction is taken into account. Nonetheless, the latter is adopted here which contributes a lower mean-field energy than the one from the single-particle states particularly at low Rabi frequencies or at relatively strong interactions~\cite{Chen2020A}. Therefore, the ground-state wave function $\Psi\left({\bf r}\right)$ is then determined from the minimization of the mean-field energy in Eq.~\eqref{eq:total-E} with respect to the variational parameters. It's worth mentioning that, three definite-QAM states are merely sufficient here, regarding the recent experimental configuration with the transferred orbital angular momentum $n=\pm1$~\cite{Zhang2019G}. One may need more definite-QAM states in the variational \textit{Ansatz} for larger $n$.
\begin{figure}[t]
\centering
\includegraphics[width=0.48\textwidth]{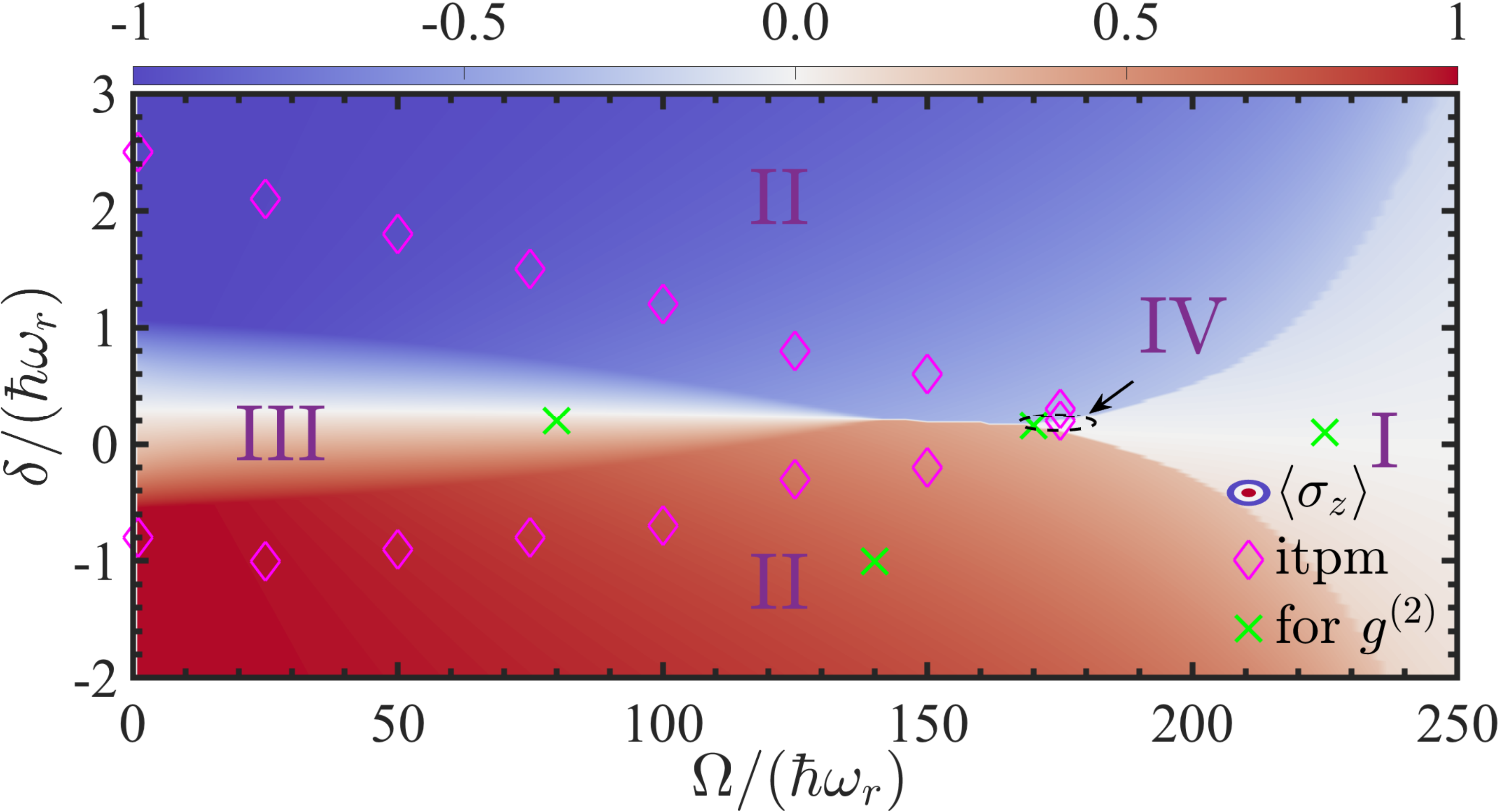}
\caption{A typical phase diagram in the $\delta$-$\Omega$ plane of a SOAM-coupled $^{41}$K gas with inter- and intra-species interactions modulated by the Feshbach resonance technique. The color indicates the expectation value of spin magnetization $\langle\sigma_z\rangle$ in the condensate wave function determined by a variational approach. The magenta diamonds indicate the regime of angular stripe phase III determined by another imaginary-time propagation method (itpm). The green crosses indicate the selected positions for four distinct phases in Fig.~\ref{fig_n_g2}, respectively. \added{Here, the Roman numbers stand for the same phases as those in Fig.~\ref{fig_n_g2}.} Adapted from Ref.~\cite{Chen2020A}.}
\label{fig_K41}
\end{figure}

\subsection{Typical ground-state phases}

\begin{table*}[ht]
\caption{\label{tab:table-phases}
A table of the classification of the typical phases in a realistic SOAM-coupled $^{87}$Rb gas with the transferred angular momentum $n =-1$ at vanishing two-photon detuning $\delta=0$~\cite{Zhang2019G}.}
\begin{ruledtabular}
\begin{tabular}{c||cccc}
&{\it Vortex-antivortex pair phase} &{\it Half-skyrmion phase} &{\it Angular stripe phases} &Reference\\
\hline\hline
Spin polarizations &$\langle\sigma_{z}\rangle=0$, $\langle\sigma_{x}\rangle=-1$ &$\langle\sigma_{z}\rangle\neq0$, $\langle\sigma_{x}\rangle\neq0$ &$\langle\sigma_{z}\rangle=0$, $\langle\sigma_{x}\rangle\neq0$ &\cite{Zhang2019G,Chen2020A,Duan2020S} \\
Magnetization &non-magnetic &magnetic &non-magnetic &\cite{Zhang2019G,Chen2020A,Duan2020S} \\
Coefficients in Eq.~\eqref{eq:2D-variationalansatz} &$\beta=1$ &$\alpha=1$ or $\gamma=1$ &$\alpha=\gamma=1/\sqrt{2}$ or $\alpha,\beta,\gamma\neq0$ &\cite{Chen2020A} \\
$l_z$ in Eq.~\eqref{eq:definite-am ansatz} &0 &$1$ or $-1$ &indefinite &\cite{Chen2020Ground,Chiu2020V,Duan2020S} \\
$\langle\hat{L}_z\rangle$ in $\psi_{_{l_{z}\uparrow\downarrow}}$ &$(1,-1)$ &$(2,0)$ or $(0,-2)$ &indefinite &\cite{Chen2020Ground,Chiu2020V,Duan2020S} \\
Symmetry\footnotemark[1] &$\mathcal{T}$ &$\cancel{\mathcal{T}}$ &$\mathcal{T}$ &\cite{Duan2020S} \\
Symmetry in $n_{\uparrow\downarrow} (n)$ &$\mathcal{R} (\mathcal{R})$ &$\mathcal{R} (\mathcal{R})$ &$\mathcal{C}_2 (\mathcal{C}_2)$ or $\cancel{\mathcal{C}_2} (\mathcal{C}_2)$ &\cite{Chen2020A}\\
Single-particle dispersion &single minimum &double degeneracy &double degeneracy &\cite{Chen2018S,Zhang2019G,Chen2020A,Chen2020Ground} \\
Excitation spectrum &symmetric, single phonon &asymmetric, roton structure &symmetric, two phonons &\cite{Chen2020A,Chen2020Ground} \\
\end{tabular}
\end{ruledtabular}
\footnotetext[1]{$\mathcal{T}$, $\mathcal{R}$ and $\mathcal{C}_2$ denote the time-reversal symmetry, continuous and two-fold rotational symmetries, respectively.}
\end{table*}

\begin{figure}[t]
\centering
\includegraphics[width=0.48\textwidth]{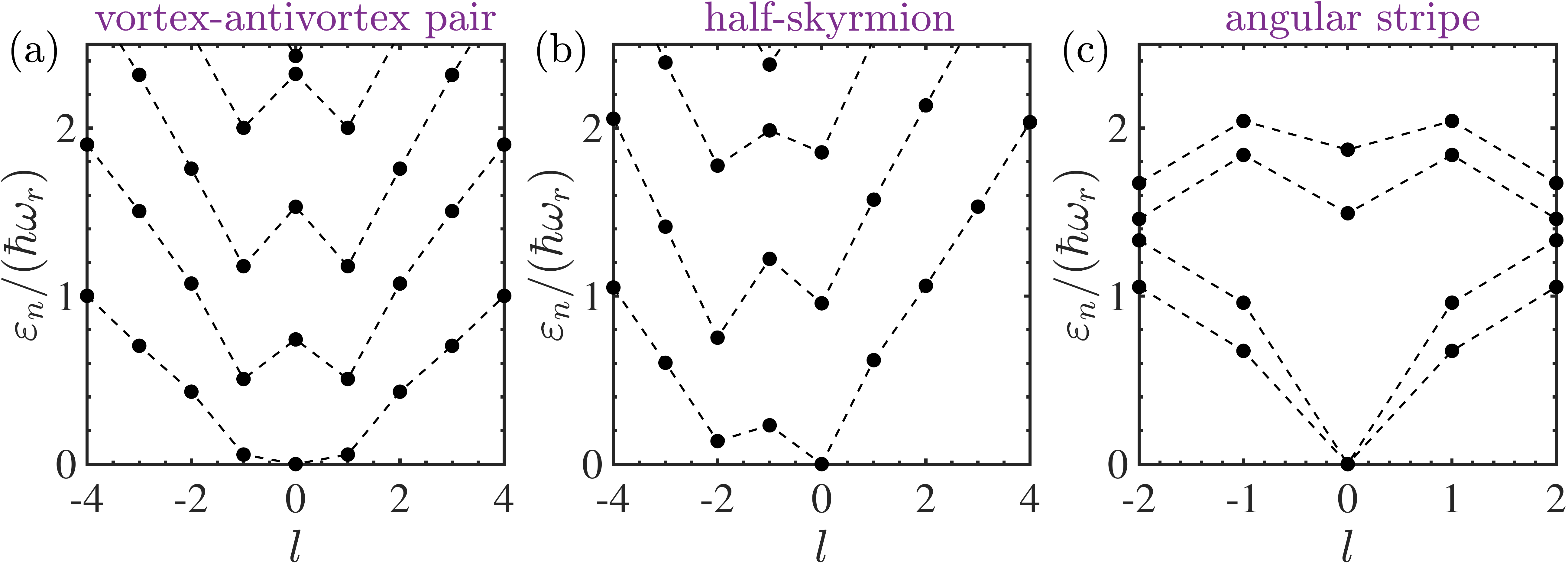}
\caption{Excitation spectrum as a function of the quasi-angular momentum $l$ of Bogoliubov quasiparticles in three typical ground-state phases, i.e., (a) vortex-antivortex pair phase, (b) half-skyrmion phase, and (c) angular stripe phase. The figures are remade based on data in Refs.~\cite{Chen2020A, Chen2020Ground}.}
\label{fig_excitation}
\end{figure}

After solving self-consistently the GP equation in Eq.~\eqref{eq:GP-eq-SOAMC} and minimizing the total energy in Eq.~\eqref{eq:total-E} with a constructed variational {\it Ansatz}, the ground-state wave function $\Psi({\bf r})$ can then be calculated and the phase diagram at zero temperature is conveniently depicted for a weakly-interacting SOAM-coupled Bose gas~\cite{Chen2020A}. In general, four distinct phases can be identified in the parameter space of $g_{\sigma\sigma'}$ and $\Omega$ at a vanishing two-photon detuning $\delta=0$. The first is the {\it vortex-antivortex pair phase} which behaves as a vortex, as shown by the first column in Fig.~\ref{fig_n_g2}. It locates at zero QAM $l_{z}=0$ (i.e., $\beta=1, \alpha=\gamma=0$) with zero magnetization $\langle\sigma_{z}\rangle=0$ and $\langle\sigma_{x}\rangle=-1$. The second one is the {\it half-skyrmion phase} with two spin-component densities being a Thomas-Fermi-like distribution and a vortex (see the second column in Fig.~\ref{fig_n_g2}). This phase is magnetic, i.e., $\langle\sigma_{z}\rangle\neq0$ and $\langle\sigma_{x}\rangle\neq0$, with a definite QAM at $l_{z}=-1$ or $1$ (i.e., $\alpha=1, \beta=\gamma=0$ or $\gamma=1, \alpha=\beta=0$). The third and fourth ones are called {\it angular stripe phases} shown by the peanut-like and halo-like density profiles in the last two columns of Fig.~\ref{fig_n_g2}. Both of them have no definite $l_{z}$ with $\alpha=\gamma=1/\sqrt{2},\beta=0$ or $\alpha=\gamma\neq0,\beta\neq0$ and they have no magnetization $\langle\sigma_{z}\rangle=0$ and $\langle\sigma_{x}\rangle\neq0$. In addition, one can further investigate the elementary excitation spectrum to distinguish these phases and then \deleted{to} characterize the phase transitions. The spectrum of a SOAM-coupled Bose gas is discrete due to the quantized angular momentum. In detail\deleted{s}, the vortex-antivortex pair phase possesses a symmetric spectrum with a single phonon mode, see Fig.~\ref{fig_excitation}(a). However, the half-skyrmion phase breaks spontaneously the double degeneracy in the single-particle dispersion and thus has an excitation spectrum with a discrete roton-maxon structure~\cite{Chen2020A}, as shown in Fig.~\ref{fig_excitation}(b). Most interestingly, owing to the spontaneous breaking of both the U$(1)$ gauge symmetry and the continuous rotational symmetry, the excitation spectrum in the angular stripe phase exhibits two gapless Goldstone modes~\cite{Chen2020Ground}, see Fig.~\ref{fig_excitation}(c). A summary of \added{the} properties of these typical phases can be found in Table~\ref{tab:table-phases}. It should be noted that these phases described above are based on recent experimental configuration\added{s} with small transferred angular momentum $n$, and we use them for a brief illustration. It is possible to find more nontrivial phases such as other angular stripe phases with superposition of large-angular-momentum states under the condition of larger $n$~\cite{Chen2020Ground,Chiu2020V} or complex vortex molecule states~\cite{Duan2020S}. Among these typical ground-state phases, the most intriguing one is the angular stripe phase, which attracts intensive research attention\deleted{s} and will be discussed in detail in the following.

\subsection{Exotic angular stripe phases}

In the past decade, an exotic phase of matter has attracted tremendous attention in ultracold quantum gases, i.e., namely the stripe phase, which breaks spontaneously U$(1)$ gauge symmetry and spatial translational symmetry~\cite{wang2010spin,Wu2011U,ho2011bose,Li2012Q}. As a consequence, quantum gases will exhibit a superfluid behaviour and meanwhile show a periodic modulation in the density distribution. These counterintuitive features associate closely this nontrivial phase to the long-sought supersolid phase in solid Helium since the 1960s~\cite{Boninsegni2012C}, i.e., a rigid, spatially ordered solid that flows like a fluid without friction. Whether such a superfluid state can exist remains unclear for more than 50 years until the seminal breakthroughs using ultracold atoms. In 2017, two groups from ETH Zurich and MIT created
successfully an ultracold quantum gas featuring supersolid properties using Bose-Einstein condensates with optical cavities or spin-orbit coupling~\cite{leonard2017supersolid,li2017stripe}. Later in 2019, three independent groups from Stuttgart, Florence, and Innsbruck observed supersolidity using dipolar quantum gases of lanthanide atoms (i.e., $^{162}$Dy, $^{164}$Dy\added{,} and $^{166}$Er) without any external optical lattice~\cite{tanzi2019observation,bottcher2019transient,chomaz2019long}, and very recently one of them
extended successfully into two dimensions~\cite{norcia2021two}.

In the presence of SOAM coupling, several kinds of angular stripe phases are theoretically predicted as introduced above, which take the analogous behavior as that of supersolid.\added{ Here, we have discussed two kinds of angular stripe phases: phase (III), a superposition state of two orbital-angular-momentum states ($l_{z}=\pm 1$), and phase (IV), a superposition state of three orbital-angular-momentum states  ($l_{z}=0,\pm 1$).} Although experimentalists have achieved successfully most of \added{the} ground-state quantum phases of a SOAM-coupled Bose gas, the nontrivial angular stripe phase remains elusive in the laboratory. There are two main obstacles. Firstly, \added{the} typical energy scale of the SOAM coupling is about $E_L\sim 1/R^2$ with the atomic \added{cloud} size $R$. \replaced{To}{In order to} enhance the SOAM-coupling effect to reach the regime of the angular stripe phase, one need\added{s} to reduce the atomic size $R$ which decreases the perimeter of the cloud and instead presents \replaced{fewer}{less} periods of density order in the angular direction. Secondly, recent experimental attempts are made in a $^{87}$Rb BEC. The critical Rabi frequency of the angular stripe phase is thus small due to the tiny difference between the intra- and inter-species interactions. As a consequence, the period of stripes becomes so small that the stripes show too low contrast and visibility, which makes them hard to be detected.

The interatomic interaction is one of the crucial factors to affect the parameter space of angular stripe phases as well as the visibility. This is quite similar to the situation when concerning the stripe phase in a SO-coupled Bose gas \cite{Li2012Q}. The stripe phase prefers the region \replaced{where}{that} the inter-species interaction is pretty smaller than the intra-species interaction. To this end, $^{41}$K atomic gases provide a promising candidate for exploring angular stripe phases with tunable interactions according to the Feshbach resonance centered at the magnetic field $B_{0}=51.95$G \cite{Chen2020A}. Near the Feshbach resonance, the intraspecies scattering lengths are approximately constant, and the interspecies one can be tuned in a wide range \cite{Lysebo2010F,Tanzi2018F}. By setting typical realistic parameters, the phase diagram is determined and the angular stripe phases can occupy a relatively large parameter space, as shown in Fig.~\ref{fig_K41}. An angular density-density correlation function $g_{i}^{(2)}(\theta)\equiv \int_{0}^{2\pi}n_{i}(\varphi)n_{i}(\varphi+\theta)\mathrm{d}\varphi/\int_{0}^{2\pi}n_{i}^{2}(\varphi)\mathrm{d}\varphi$ can be introduced to estimate the visibility in density profile, with the angular density $n_{i}(\varphi)=\int_{0}^{\infty}r\mathrm{d}rn_{i}(r,\varphi)$, and the label $i=\uparrow,\downarrow$ for each spin component and null for the total density. In contrast to other phases, the novel angular stripe phases break the rotational symmetry and exhibit spatial density modulation in the angular direction. Two angular stripe phases feature considerable spatial modulation and contrast in the angular density-density correlation for both spin components as well as the total one, as shown in Figs.~\ref{fig_n_g2}(c) and \added{\ref{fig_n_g2}}(d). This hallmark feature might be useful in directly probing the existence of the angular stripes in future experiments with ultracold atoms.

There are also other theoretical attempts and developments emphasizing \deleted{on} the angular stripe phases in SOAM-coupled Bose gases. For example,
by utilizing LG beams with higher-order orbital angular momenta, it is shown that angular stripe phases can be achieved in a wide window of experimentally accessible parameters with high visibility contrast \cite{Chen2020G,Chiu2020V}. Besides, A number of distinct quantum phases are identified according to the symmetry analysis, and a complex vortex molecule state is discovered, which plays an important role in the continuous phase transitions \cite{Duan2020S}. For a ring-shaped BEC in the presence of SOAM coupling, the fine structures of angular stripe phases are explored \cite{Bidasyuk2022F}.

\section{SOAM-coupled Fermi gases}
\label{sec:SOAMCFermi}
The pairing mechanism plays a crucial role \replaced{in}{on} the Fermi superfluid. While it is shown that the SOAM coupling leads to the spin-dependent vortex formation in Bose\added{-Einstein} condensates, SOAM coupling alone does not induce vortices in a Fermi superfluid, since fermions in a Cooper pair would acquire opposite orbital angular momenta that cancel each other, yielding a superfluid devoid of vortices. In this section, we are going to discuss \added{the} unique features of SOAM-coupled Fermi superfluid at zero temperature, with emphasis on two exotic pairing states, i.e., SOAM-coupling-induced vortex and topological superfluid states.

\subsection{Pairing physics under SOAM coupling}
\label{sec:soamc}

It is well known that a two-component spin-balance Fermi gas becomes unstable in the presence of an arbitrary small attractive interaction. The resulting instability gives rise to a Bardeen-Cooper-Schrieffer (BCS) ground state with zero center-of-mass momentum.  It is also well accepted that in a Fermi superfluid with spin-imbalance or Zeeman fields, exotic pairing states emerge,  for instance, Fulde-Ferrell  and Larkin-Ovchinnikov states~\cite{FF, LO}, where the former pairing state carries finite center-of-mass momentum and the latter oscillates both in coordinate and momentum spaces. Whereas,  these novel states are only stable in a very small parameter region, which is one of the reasons for the exclusiveness of their experimental observations. However,  the experimental achievement of SO coupling in cold atoms offers a new platform to pursue these long sought-after states. For instance,  the interplay of SO coupling, Zeeman fields and interactions provides an alternative mechanism to induce Fulde-Ferrell-Larkin-Ovchinnikov (FFLO) states~\cite{Dong2013F,Shenoy2013F,Wu2013U,Qu2013T,Zhang2013T,Chen2013I,Liu2013T}.   Besides SO-coupling-induced FFLO states,  topological band structures become accessible under SO coupling~\cite{Huang2016E, Wu2016R, Zhang-16, xjliu1D, xjliu2D},  which lays  the fertile ground for topological superfluid--for instance, aided by the Zeeman fields and interactions, a topological superfluid emerges in a 2D Fermi gas under the Rashba-type SO coupling~\cite{Kane-08,tsfsolid1,tsfsolid2,tsfsolid3,tsfsolid4,soc1,Yi-11}.

Intuitively,  SOAM coupling  as \replaced{an}{the} angular analog of the conventional SO coupling,  two novel pairing states can be expected. One is the SOAM-coupling-induced pairing state with finite quantized angular momenta,  which is nothing but a vortex state.  The other is the SOAM-coupling-induced angular topological superfluid state, the analog of the topological superfluid state in a one-dimensional lattice gas under a one-dimensional SO coupling, where quantized angular momenta in the former play the role of discretized linear momenta of the latter.  In this review,  whether the above two expectational pairing states, i.e., SOAM-coupling-induced vortex and topological superfluid states,  could be stabilized is clarified. In the following, the former issue will be addressed in Sec.~\ref{sec:vortex} and the latter in Sec.~\ref{sec:TSF}.

\begin{figure}[t]
\begin{center}
\includegraphics[width=0.48\textwidth]{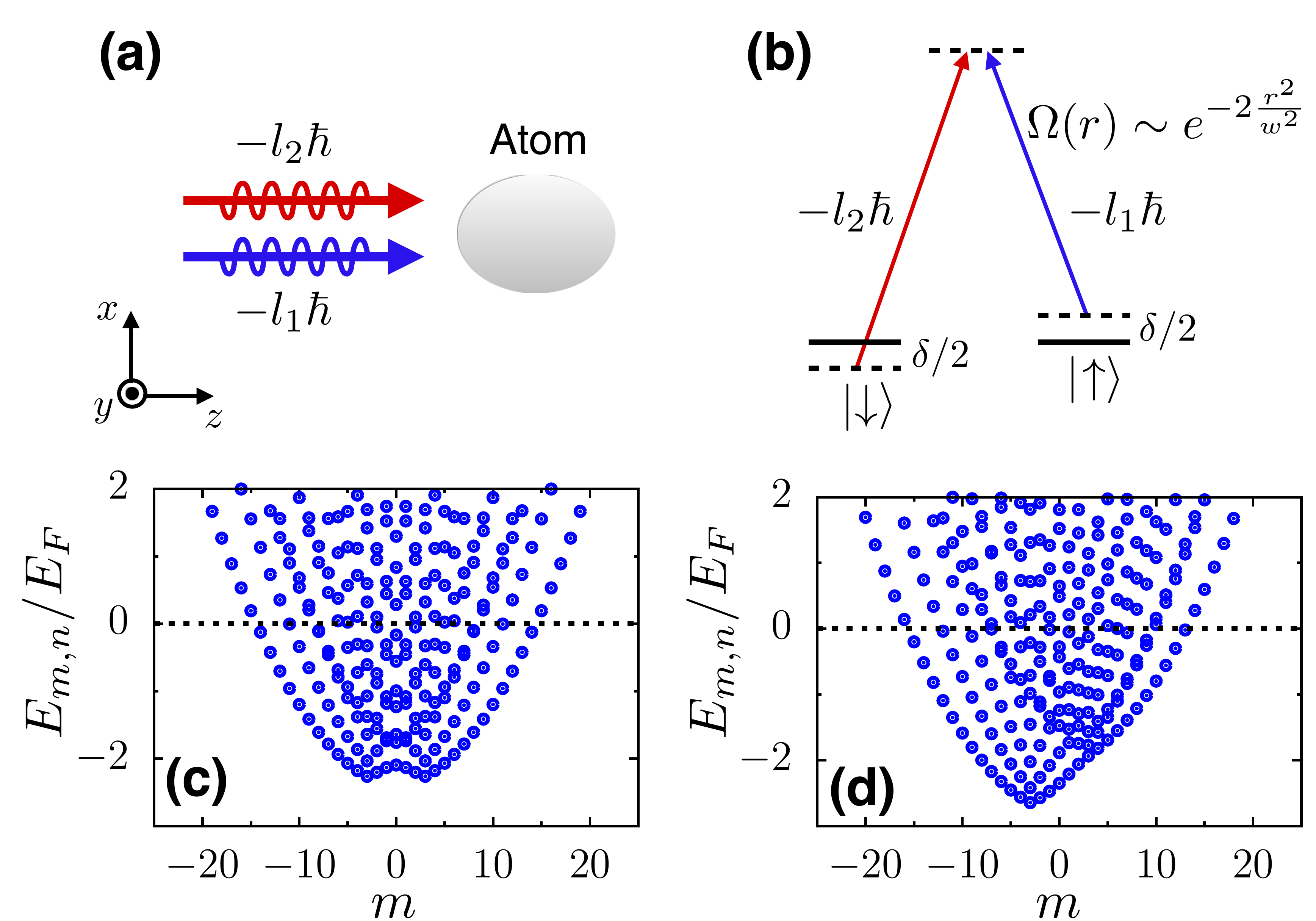}
\caption{(a) SOAM coupling in atoms induced by a pair of copropagating Raman beams carrying different orbital angular momenta ($-l_1\hbar$ and $-l_2\hbar$), with a transferred angular momentum $2l\hbar=(l_1-l_2)\hbar$. (b) Schematic illustration of the level scheme.
(c)(d) Single-particle energy spectra under SOAM coupling for $\delta=0$ (c), and $\delta /E_F= 0.4$ (d) with $E_F=\hbar^2 k^2_F/(2M)$  the Fermi energy and $k_F$ the Fermi vector. Black dashed lines denote potential Fermi surfaces in a many-body setting.
Adapted from Ref.~\cite{Chen2020G}.}
\label{Fig1-Fermi}
\end{center}
\end{figure}

\subsection{SOAM-coupling-induced vortex states}
\label{sec:vortex}
As illustrated in Fig.~\ref{Fig1-Fermi},  a two-component Fermi gas is confined in the $x-y$ plane, where the  two ground states are labeled by $\uparrow$ and $\downarrow$,  respectively.  The two-photon Raman process is driven by two copropagating Raman beams carrying different orbital angular momenta $-l_1\hbar$ and $-l_2\hbar$ [see Figs.~\ref{Fig1-Fermi}(a),~\ref{Fig1-Fermi}(b)], and is characterized by the inhomogeneous Raman coupling $\Omega(r)$, two-photon detuning $\delta$ and a phase winding $e^{-2il\varphi}$. Here,  the polar coordinate  ${\bf r}=(r,\varphi)$ is taken.  After a unitary transformation,  the effective single-particle Hamiltonian again reduces to Eq.~\eqref{eq:1Ham2}. As we can see that the Raman coupling $\Omega(r)$ and two-photon detuning $\delta$
serve as effective transverse and longitudinal Zeeman fields, respectively, which play crucial roles in stabilizing vortices.   Different from the LG Raman beams used in  previous experiments ~\cite{Chen2018S, Zhang2019G}, Chen {\it et al}. propose Raman coupling as $\Omega(r)=\Omega_0 e^{-2r^2/w^2}$~\cite{Chen2020G},  with $\Omega_0$ the peak intensity and $w$ the beam waist, which can be experimentally realized~\cite{Rumala-17}.  Such a \replaced{choice}{choose} of the Raman beams is mainly on the basis that the LG Raman beams used in current experiments are suppressed over a considerable region near $r=0$,  giving rise to \added{an} almost vanishing spin mixing effect in the vicinity, which is unfavorable for vortex formation essentially. Here, Raman lasers are assumed to  operate at the tune-out wavelength~\cite{Holmgren2012M,Herold2012P,Zhang2019G}, leading to  a vanishing diagonal ac-stark potential that is consistent with the experiment \cite{Zhang2019G}.  The external potential $V_{ext}(r)$ is chosen as an isotropic hard-wall box trap with a radius $R$, which offers a natural boundary.

It is helpful for understanding the mechanism of vortex formation by analysis of the single-particle properties before moving to the many-body calculation.  Figs.~\ref{Fig1-Fermi}(c) and (d)  illustrate the impact of the two-photon detuning $\delta$ on the single-particle spectrum. At the resonance with $\delta=0$, the time-reversal symmetry gives rise to $E_{m,n}=E_{-m,n}$. Whereas away from the resonance at finite $\delta$, the breaking of the time-reversal symmetry results in a deformation of the Fermi surface. In the many-body setting when the attractive interaction is taken into account, pairing predominantly occurs between unlike spins with the same radial quantum number $n$ to maximize the overlap of radial wave functions.
Thus, for the symmetric eigenspectrum under $\delta=0$, it is more favorable for two fermions with opposite angular quantum numbers ($m$ and $-m$), forming  a Cooper pair with a zero total angular momenta.
In contrast, under a finite $\delta$ with asymmetric eigenspectrum, the two fermions in a Cooper pair may possess different values of $|m|$, leading to a pairing state with a nonzero quantized angular \replaced{momentum}{momenta}, which is the so-called vortex state. Such a mechanism for the vortex formation is analogous to that of the SO-coupling-induced Fulde-Ferrell states, where the interplay between SO coupling and Zeeman fields gives rise to the deformation of Fermi surfaces with broken time-reversal symmetry in the momentum space~\cite{Wu2013U,Qu2013T,Zhang2013T,Chen2013I,Liu2013T}.

\begin{figure}[tbp]
\begin{center}
\includegraphics[width=0.45\textwidth]{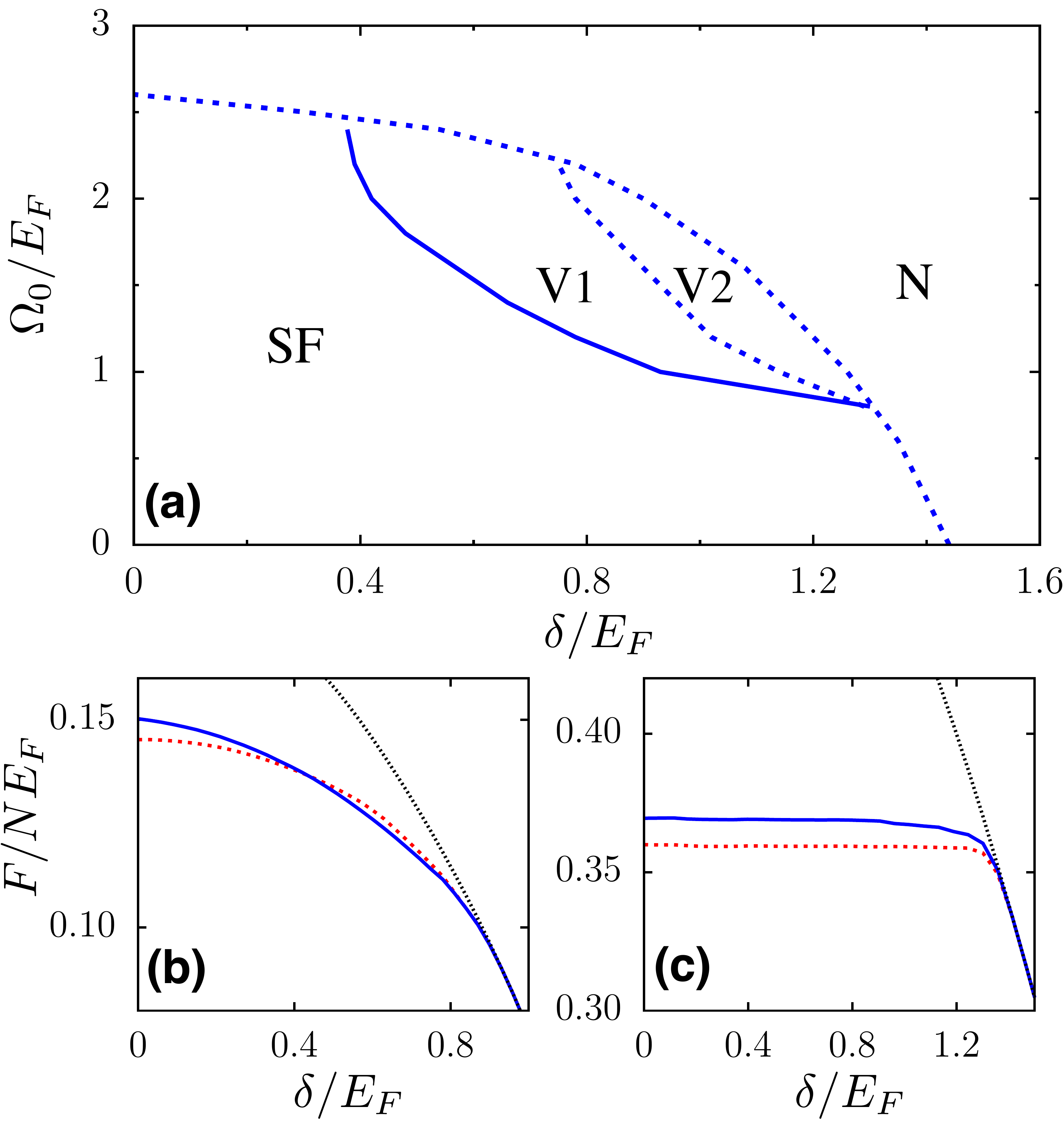}
\caption{(a) Phase diagram of a two-dimensional Fermi superfluid with SOAM coupling in the $\Omega_0$-$\delta$ plane. The phase diagram includes the usual superfluid state  with $\kappa=0$, the normal state  with $\Delta=0$, and two vortex states with $\kappa=-1$: a fully gapped vortex states ($V1$) and a gapless vortex state ($V2$).
(b)(c) Free energies $F$ of the superfluid ($\kappa=0$; red dashed), vortex ($\kappa=-1$; blue solid) and normal (black dotted) states as functions of $\delta$, with $\Omega_0/E_F=2$ (b) and $\Omega_0/E_F=0.5$ (c).
Adapted from Ref.~\cite{Chen2020G}.}
\label{Fig2-Fermi}
\end{center}
\end{figure}

Chen {\it et al}. confirm the above analysis by solving the many-body problem under the Bogoliubov--de Gennes (BdG) formalism~\cite{Chen2020G}.  Specifically, the order parameter is written in the form of $\Delta({\bf r})=\Delta(r)e^{i \kappa \varphi}$, where the vorticity $\kappa=0$ ($\kappa \neq 0$)  indicates the superfluid (vortex) state. With different $\kappa \in \mathbb{Z}$,   the BdG equation and $\Delta({\bf r})$ are solved self-consistently with \added{a} fixed particle number. The ground state is then  determined by comparing  the free energies of vortex states with different $\kappa$, the usual superfluid state ($\kappa=0$), and the normal state ($\Delta=0$).

As shown in Fig.~\ref{Fig2-Fermi}(a),  Chen {\it et al}.  give \added{a} typical phase diagram of the system with SOAM coupling in the $\Omega_0$--$\delta$ plane.  For $\delta>0$ ($\delta<0$),  vortex states with $\kappa=-1$ ($\kappa=1$) emerge, with the phase boundaries  unchanged to the sign of $\delta$. At small $\Omega_0$ and $\delta$,   the ground state is a usual superfluid (SF) with a zero vorticity $\kappa=0$.  Under sufficiently large $\Omega_0$ and/or $\delta$, the free-energy difference between the SF and normal ($N$) states becomes vanishingly small, and hence the system enters the normal state. Remarkably, two vortex states emerge between SF and $N$ states. For example, with a fixed $\Omega_0/E_F=1.5$ [see Fig.~\ref{Fig2-Fermi}(a)], the ground state is in the SF state under small detunings $\delta$, and becomes a fully-gapped vortex state ($V1$) beyond a critical value of $\delta$. Further increase of $\delta$ gives rise to a gapless vortex state ($V2$),  whose bulk excitation gap closes.  In Figs.~\ref{Fig2-Fermi}(b),~\ref{Fig2-Fermi}(c), free energies of different states are compared, as the phase diagram is traversed. Especially, for the case with $\Omega_0=0.5E_F$, the ground state remains vortexless for finite $\delta$, despite the deformation of the Fermi surface under SOAM coupling and effective Zeeman fields.  This originates from  the quantized nature of the angular momentum, and is on the sharp contrary to the SO-coupling-induced Fulde-Ferrell state which carries a nonzero, continuously varying center-of-mass momentum in the presence of SO coupling and Zeeman fields~\cite{Dong2013F,Shenoy2013F,Wu2013U,Qu2013T,Zhang2013T,Chen2013I,Liu2013T}.

\begin{figure}[tbp]
\begin{center}
\includegraphics[width=0.4\textwidth]{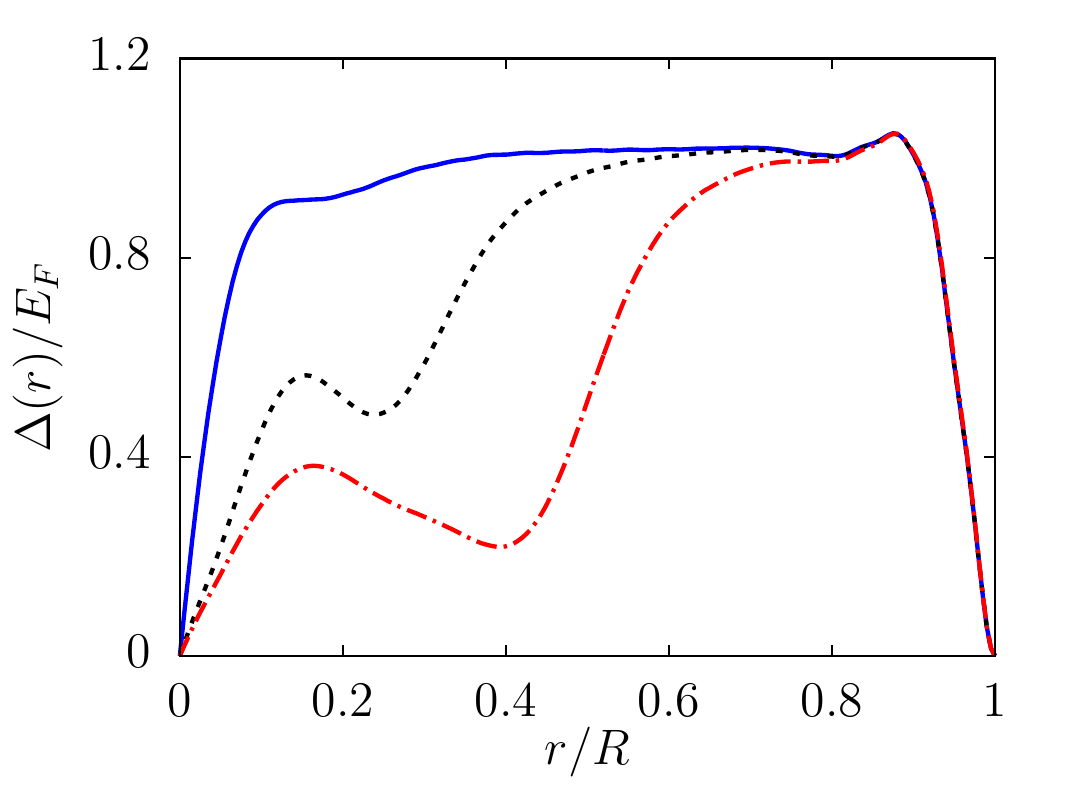}
\caption{Order parameter profiles for $k_Fw=5$ (blue solid), $k_Fw=10$ (black dashed), and $k_Fw=15$ (red dash-dotted). We fix $\delta/E_F=0.84$ and $k_FR=15$. Adapted from Ref.~\cite{Chen2020G}.}
\label{Figvortex}
\end{center}
\end{figure}

Remarkably,  as illustrated in Fig.~\ref{Figvortex},  the SOAM-induced vortex state features a giant and tunable vortex-core size, \added{similar to that in a BEC interacting with a microwave field~\cite{Qin2016S}}. The vortex-core size, characterized by variations of the order parameter, is comparable to the waist $w$ of LG beams. Compared to the conventional vortex states in atomic Fermi superfluids, where  changes in the vortex-core structure predominantly take place within a short length scale set by the interatomic separation~\cite{Sensarma2006V,Chien2006G}, these SOAM-coupling-induced vortex states, with tunable size and core structure,  provide unprecedented experimental access to topological defects in Fermi superfluids.

Different from Chen {\it et al}.'s configuration, Wang {\it et al.} investigate a similar vortex-forming scheme under SOAM coupling~\cite{Wang2021E}. Most strikingly, they predict that an unprecedented vortex state, which is an angular analog of SO-coupling-induced Larkin-Ovchinnikov state, to occur.

Nevertheless, for the inevitable heating introduced by the Raman  process, it is difficult to cool a realistic Fermi gas with SOAM coupling below the superfluid temperature~\cite{heating},  which  leads to the exclusiveness of  SOAM-coupling-induced novel states. Instead, concerning the  persistence of dressed molecules above the critical temperature in Fermi gases with SO coupling~\cite{Williams2013R,Fu2014P}, it is reasonable to expect that molecular states in \replaced{a}{an} SOAM-coupled Fermi gas should be readily accessible under typical experimental conditions.  Based on the above consideration,  Han {\it et al}.  \replaced{studied}{study} the two-body bound states in a SOAM-coupled quantum gas of fermions very recently~\cite{Han2022M}.  They identify the condition for the emergence of molecular states with finite total angular momenta  and  propose to detect the molecules according to the radio-frequency spectroscopy. As the molecular states can form above the superfluid \added{transition} temperature, Han {\it et al}. offer an experimentally more accessible route toward the study of the underlying pairing mechanism under SOAM coupling.

\subsection{SOAM-coupling-induced topological states}
\label{sec:TSF}
As discussed in Sec.~\ref{sec:soamc}, it is expected that SOAM coupling can induce a topological superfluid with the help of Zeeman fields and interactions.  While,  \replaced{implemention}{implementing} of SOAM coupling relies on the spatial dependence of the LG beams,  which gives rise to the dimensionality of atomic gases must higher than one; whereas, it is also realized that the Fermi superfluid becomes gapless and losses its topological features when its spatial dimensionality higher than that of SO coupling~\cite{Wu2013U,yisoc}.  Such that the one-dimensional nature of the SOAM coupling (coupling only occurs along the azimuthal direction)  imposes a stringent constraint on the stability of an angular topological superfluid. Whether such a topological superfluid can also be stabilized under SOAM coupling should be verified.  It is shown that \added{the stability of} a fully gapped angular topological superfluid survives the constraint above, provided that the radial motion of atoms is sufficiently suppressed and then the topological gap is not closed~\cite{Chen-22}.

\begin{figure}[tbp]
\begin{center}
\includegraphics[width=0.48\textwidth]{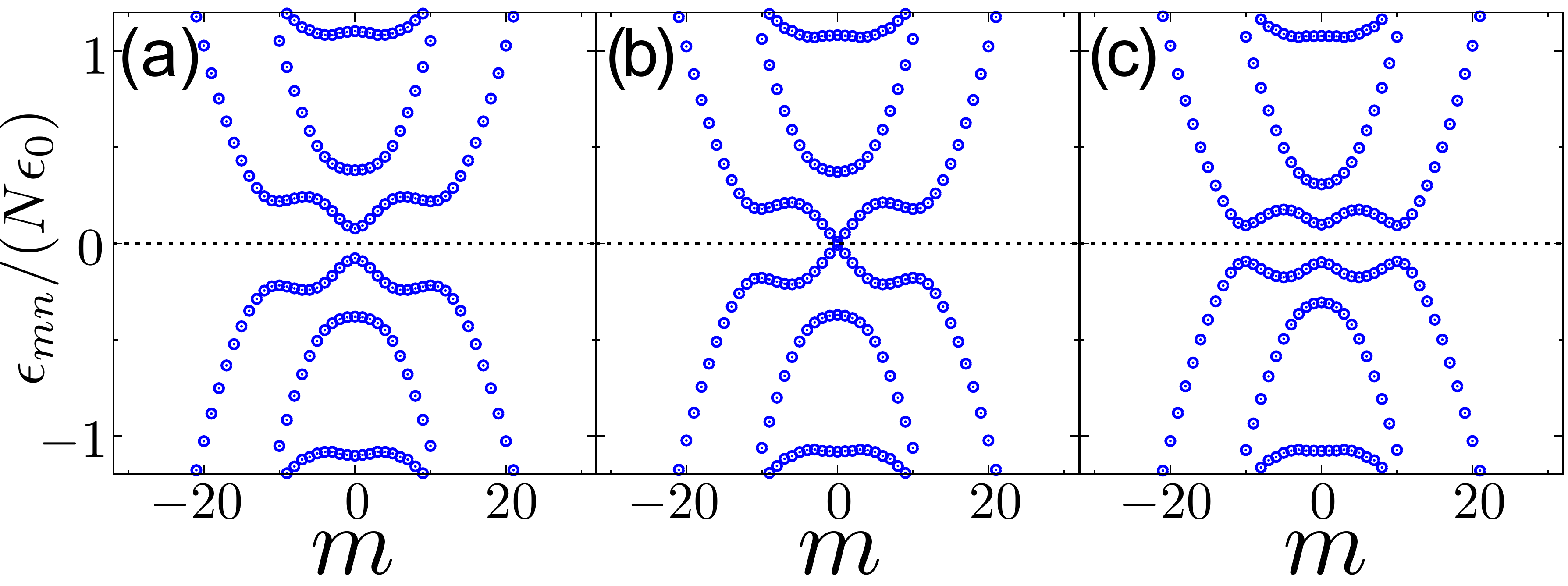}
\caption{Bogoliubov spectra of the Fermi superfluid under SOAM coupling, with a vanishing two-photon detuning $\delta=0$, and an increasing $\Omega_0$: (a) $\Omega_0/\epsilon_0=0.15$, (b) $\Omega_0/\epsilon_0=0.18$, and (c) $\Omega_0/\epsilon_0=0.2$.  Adapted from Ref.~\cite{Chen-22}. }
\label{Fig2}
\end{center}
\end{figure}

The survives of the angular topological superfluid can be understood through the following  analysis. As shown in Eq.~\eqref{eq:1Ham2}, the Raman coupling and the diagonal ac-stark potential can be written as $\Omega(r)=\Omega_0 I(r)$ and $\chi(r)=\chi_0 I(r)$, consistent with configurations of current experiments on SOAM coupling~\cite{Chen2018S, Zhang2019G}. Here, $\Omega_0$ is the effective SOAM-coupling strength, $\chi_0$ is the trapping strength of the diagonal ac-stark potential, and $I(r)= (\sqrt{2}r/w)^{2l}e^{-2 r^2 /w^2}$ is the spatial intensity profile of LG lasers, with $w$ the beam waist.  Different from \added{the} SOAM-induced vortex state in Sec.~\ref{sec:vortex}, here,  $\chi(r)$ provides an extra confinement potential, which plays an important role in stabilizing \added{the} topological superfluid. For a confinement that is sufficiently tight along the radial direction, the radial degrees of freedom of the atoms \replaced{are}{is} frozen, and the remaining quantized angular motion is then well-captured by an effective one-dimensional model with discretized modes.  Intuitively, such a scenario occurs when the trap depth $\chi_0$ is so large that the radial excitation energy ($\sim \sqrt{ |\chi_0| \hbar^2 /(m w^2)}$) becomes much larger than any other relevant energy scales of the system. The atoms are then localized near $r_0=\sqrt{l/2}w$ in the radial direction and the system reduces to an effective one-dimensional model along the angular direction.

The above analysis is confirmed through the BdG approach~\cite{Chen-22}.  As illustrated in Fig.~\ref{Fig2}, Chen {\it et al}. show the Bogoliubov quasiparticle spectrum in a sufficiently deep ac-stark potential with $\chi_0/\epsilon_0=-8$ in the unit of energy  $\epsilon_0=\pi^2 \hbar^2/(2M r^2_0)$.
For the case with $\delta=0$,  the ground state always emerges at $\kappa=0$. Here, consistent with Sec.~\ref{sec:vortex}, the vorticity $\kappa=0$ ($\kappa \neq 0$)  denotes the superfluid (vortex) state.
By increasing the coupling strength $\Omega_0$, the Bogoliubov spectrum undergoes a gap\added{-}closing and re-opening process, reminiscent of that of a topological phase transition. Specifically, the Bogoliubov quasiparticle excitation is fully gapped under small $\Omega_0$ [Fig.~\ref{Fig2}(a)], and then becomes gapless at a critical $\Omega^{c}_0/\epsilon_0\approx 0.18 $ [Fig.~\ref{Fig2}(b)], and finally is again fully gapped for further increasing $\Omega_0$ [Fig.~\ref{Fig2}(c)]. The process of gap closing and re-opening is further demonstrated as  a topological phase transition by  calculating the Zak phase  of \added{the} one-dimensional effective Hamiltonian~\cite{Chen-22}. Besides, it is also demonstrated that the angular topological superfluid is stabilized when \added{the} ac-stark potential becomes sufficiently large.

\begin{figure}[t]
\begin{center}
\includegraphics[width=0.48\textwidth]{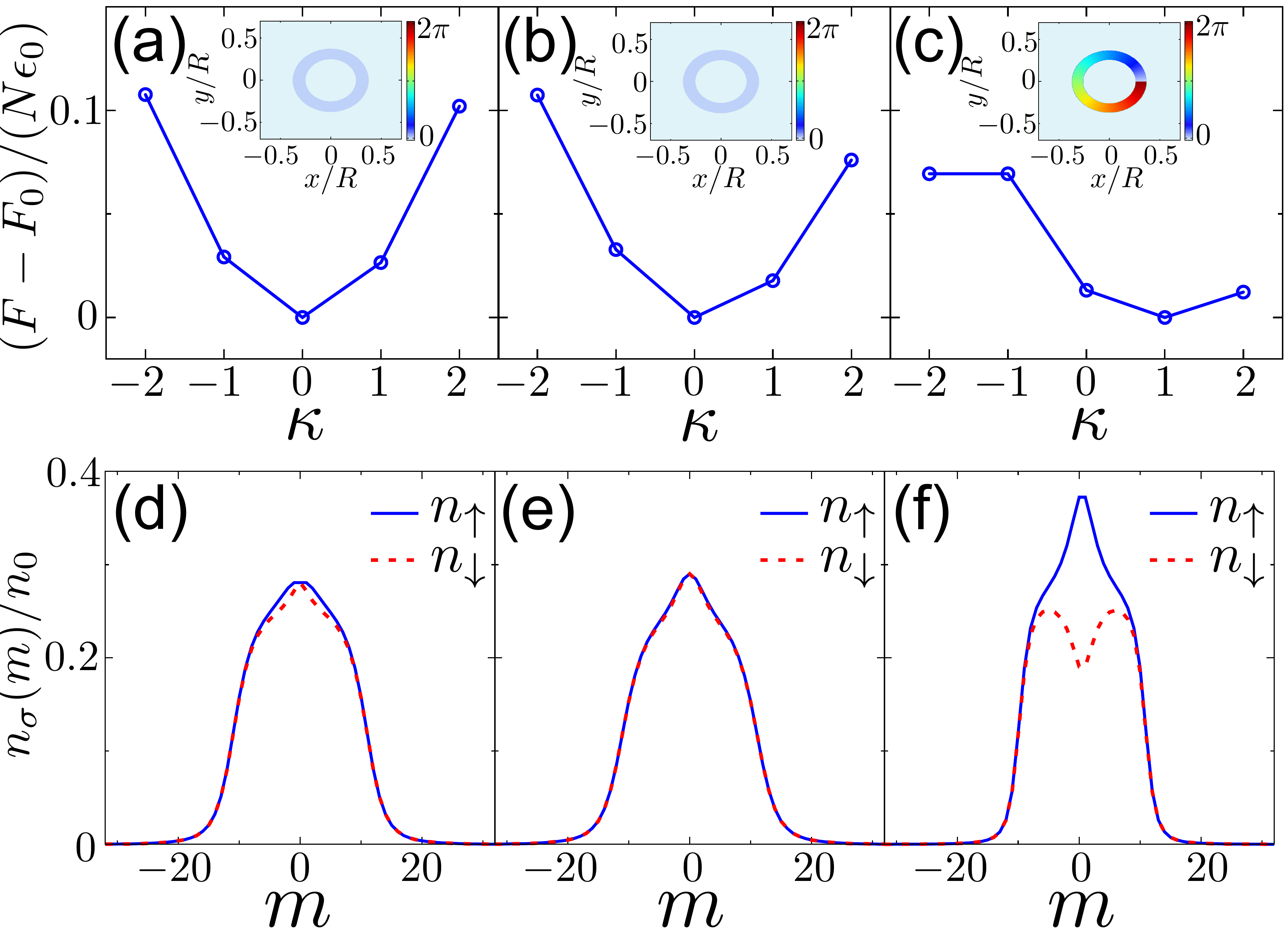}
\caption{(a)--(c) Free energies of pairing states as functions of $\kappa$, with a fixed two-photon detuning $\delta/\epsilon_0=-1.6$. The insets show the phase of the order parameter. Here, $F_0$ denotes the ground-state free energy. (d)--(f) Density distribution of the ground state in the angular-momentum space, where we define $n_0=N/(\pi r^2_0)$. Here, the blue solid (red dashed) curve denotes the density distribution of spin-up (spin-down) component. (a), (d) and (b), (e) are the vortexless superfluid states with $\Omega_0/\epsilon_0=0.1$ and $0.16$, respectively. (c), (f) is a topological vortex state with $\Omega_0/\epsilon_0=0.18$.
Adapted from Ref.~\cite{Chen-22}.}
\label{Fig4}
\end{center}
\end{figure}

Building upon the topological superfluid state above, an exotic topological vortex state can be induced by taking the two-photon detuning $\delta$ into account,  which deforms the Fermi surface. As shown in Figs.~\ref{Fig4}(a)--\ref{Fig4}(c),  the free energy is generically asymmetric with respect to $\kappa=0$ at a finite $\delta$. The asymmetry becomes more apparent with increasing $\Omega_0$, until the ground-state order parameter eventually acquires a finite phase with $\kappa\neq 0$.
Intriguingly, such a transition into the vortex state is topological. As demonstrated in \cite{Chen-22}, while the Fermi-surface deformation is manifested as the asymmetric spectral shape with respect to $m=0$, the closing and re-opening of the energy gap persist\deleted{s}. Indeed, after the gap is reopened, the angular momentum of the ground state jumps from $\kappa=0$ to $\kappa=1$. The ground state simultaneously becomes topological, which is confirmed by the Zak-phase calculation.
Conceptually, such an exotic topological vortex state is the angular version of the topological Fulde-Ferrell state under the conventional SO coupling~\cite{Qu2013T, Zhang2013T, Chen2013I, Liu2013T}.

In addition, the topological vortex leaves a direct signature in the angular-momentum-space density profile, as illustrated in Figs.~\ref{Fig4}(d)--(f). The density profile of the minority spin species exhibits a dip close to $\kappa/2$ only in the topological vortex state \replaced{[}{(}Fig.~\ref{Fig4}(f)\replaced{]}{)}, since the spin polarization in the vortex state is a direct result of the two-photon detuning, which plays the role of an effective Zeeman field. Similar signatures have been identified in the topological Fulde-Ferrell state under SO coupling.

\added{The origin of angular topological superfluid comes from the interplay of interactions, Zeeman fields and SOAM couplings. As topological superfluids are believed  to host Majorana zero modes at boundaries, once a boundary is created, for instance, by shining a strong laser beam to break the ring geometry of the ac stark potential, Majorana zero modes should be observed.}

\section{Experiment achievements}
\label{sec:Experiments}
The SOAM coupling has been achieved in $^{87}$Rb Bose gases independently
by two experiment groups~\cite{Chen2018S,Chen2018R,Zhang2019G},
following the earlier theoretical proposals~\cite{DeMarco2015A,Sun2015S,Qu2015Q}.
The Zeeman sublevels of the ground hyperfine manifold |F = 1\textrangle{} are coupled by a pair of copropagating LG beams according to a Raman transition. It leads to an orbital angular momentum change of atoms
during the transition between the ground Zeeman states, which play
the role of spin. The realization of SOAM coupling in cold atoms has recently been reported in a spin-$1$
$^{87}$Rb BEC~\cite{Chen2018S}, in which the atoms are loaded into
the middle-energy state of three Raman-dressed states. The correlations
between spin and orbital angular momentum in this Raman-dressed state
are confirmed. Then the ground-state quantum phase transitions are
observed in the same bosonic system~\cite{Chen2018R}, known as the
Hess-Fairbank effect~\cite{Hess1967M,Ishiguro2004V}. Meanwhile,
the SOAM coupling was also demonstrated in an effective spin-half
atomic gas~\cite{Zhang2019G}, following a similar Raman scheme.
The ground-state phase diagram is comprehensively studied, and the
first-order phase transitions are identified. In the follows, we are
going to introduce the main achievements in these experiments.

\begin{figure}
\includegraphics[width=0.6\columnwidth]{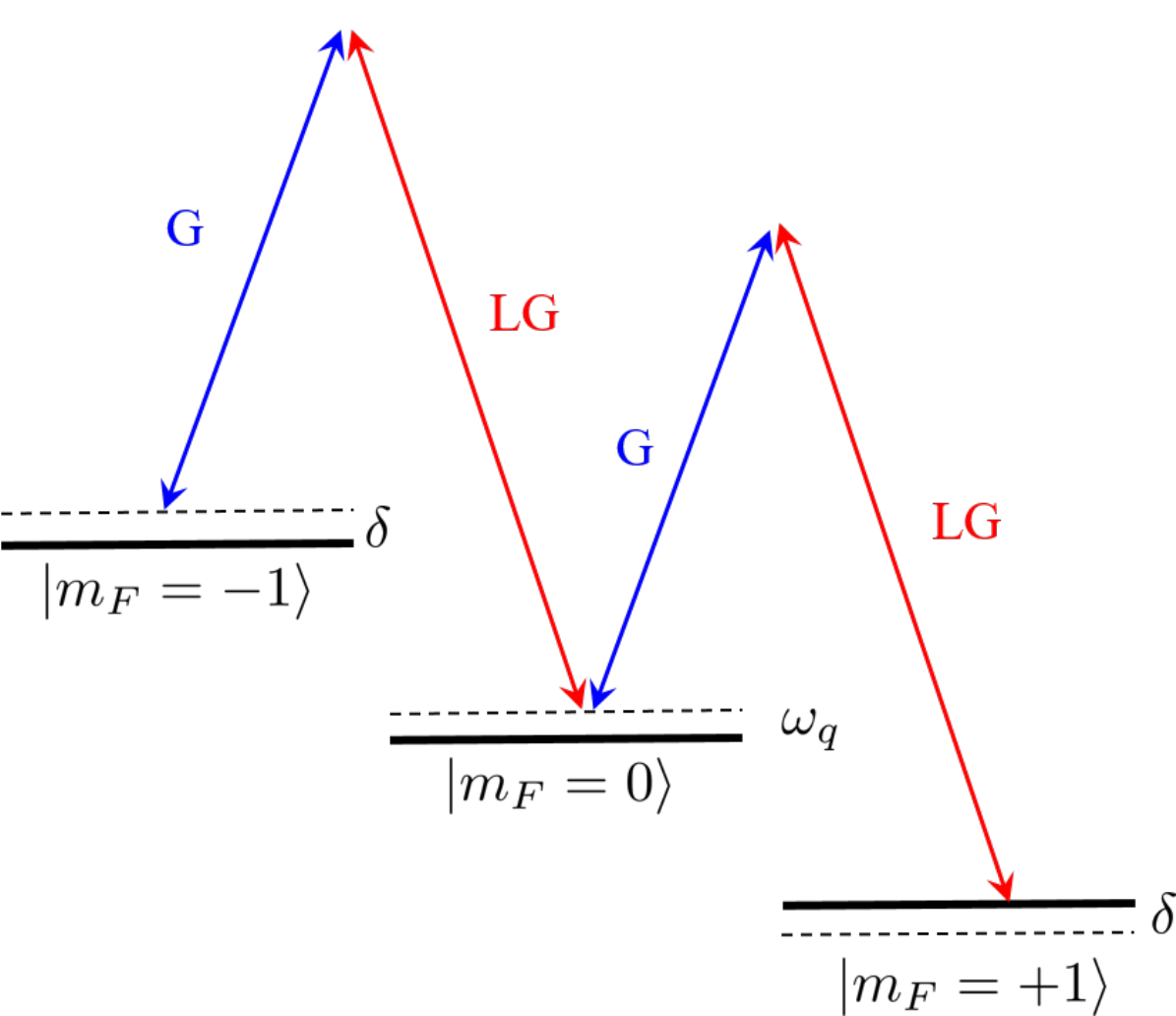}

\caption{The schematic of energy-level diagram of Raman transition in~\cite{Chen2018S}.
The hyperfine states of the ground-state manifold $\left|F=1\right\rangle $
of $^{87}$Rb atoms are coupled by a pair of Raman beams, in which
one is a LG beam denoted by ``LG'' and the other one is a Gaussian
beam denoted by ``G''. Here, $\delta$ is the Raman detuning that
is tunable in the experiment, and $\omega_{q}\approx2\pi\times50$Hz
is the quadratic Zeeman shift. }

\label{fig:LinExp1}
\end{figure}

In the experiment of Lin's group~\cite{Chen2018S}, the three hyperfine
states of the ground-state manifold $\left|F=1,m_{F}=0,\pm1\right\rangle $
of $^{87}$Rb atoms are coupled by a pair of Raman beams as illustrated
in Fig.~\ref{fig:LinExp1}, one of which is an LG beam carrying orbital
angular momentum $l=1$. The center-of-mass angular momentum of atoms
changes during the transition between different hyperfine states.
This effectively produces a light-induced effect of a spin $\left|{\bf F}\right|=1$
particle moving in a magnetic Zeeman field $\boldsymbol{\Omega}_{eff}$,
which results in a SOAM coupling after a local spin rotation. In the
experiment, the BEC is initially prepared in the bare hyperfine state
$\left|m_{F}=-1\right\rangle $, and then is transferred to $\left|m_{F}=0\right\rangle $.
After slowly switching on the Raman fields, the atoms are adiabatically
loaded into the Raman-dressed polar state $\left|\xi_{0}\right\rangle $
with $\left\langle {\bf F}\right\rangle =0$, the middle-energy eigenstate
of $\boldsymbol{\Omega}_{eff}\cdot{\bf F}$ with the form of~\cite{Ho1998S}
\begin{equation}
\left|\xi_{0}\right\rangle =-\frac{e^{i\varphi}\sin\beta}{\sqrt{2}}\left|-1\right\rangle +\cos\beta\left|0\right\rangle +\frac{e^{-i\varphi}\sin\beta}{\sqrt{2}}\left|+1\right\rangle \label{eq:RamDress0}
\end{equation}
in the bare hyperfine basis $\left|m_{F}=0,\pm1\right\rangle $, where
$\beta\left(r\right)=\arctan\left[\Omega\left(r\right)/\delta\right]$
is the polar angle of the light-induced magnetic Zeeman field $\boldsymbol{\Omega}_{eff}$.
Here, $\Omega\left(r\right)$ is the Raman-coupling strength and $\delta$
is the Raman detuning. After the adiabatic loading, one easily find\added{s}
a correlation between the atom spin (hyperfine state) and its orbital
angular momentum: the atoms acquire an orbital angular momentum $\Delta l=1$
($\Delta l=-1$) when transitioning from $\left|m_{F}=0\right\rangle $
to $\left|m_{F}=-1\right\rangle $ ($\left|m_{F}=+1\right\rangle $).
Therefore, the vortex structures are anticipated in the bare hyperfine
states $\left|m_{F}=\pm1\right\rangle $. The Raman-dressed state
$\left|\xi_{0}\right\rangle $ is conveniently characterized by the
QAM $l_{z}=0$ in the rotating frame, and the angular momenta of bare
hyperfine states in the laboratory frame are, respectively, $l_{m_{F}=0}=l_{z}$
and $l_{m_{F}=\pm1}=l_{z}\mp1$. 

\begin{figure}
\includegraphics[width=1\columnwidth]{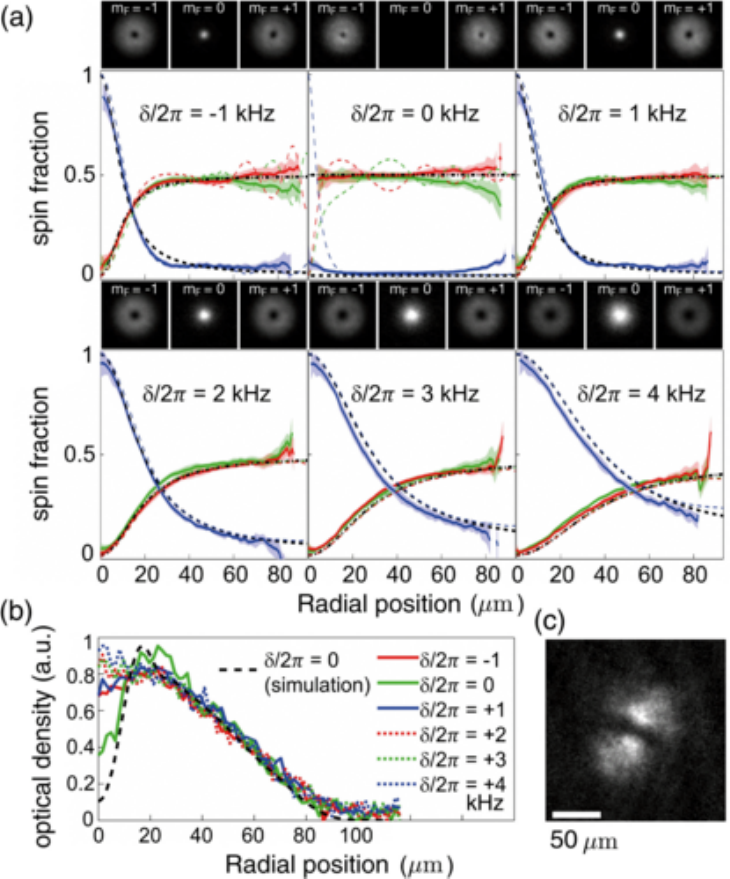}

\caption{Demonstration of SOAM coupling in $^{87}$Rb BEC adiabatically loaded
into the Raman-dressed state $\left|\xi_{0}\right\rangle $. The images
are taken after $24$ms TOF with holding time $1$ms. (a) The density
profiles of bare hyperfine components $\left|m_{F}\right\rangle $
for different Raman detunings. The blue, red, and green curves are
experimental data for $\left|0\right\rangle $, $\left|+1\right\rangle $
and $\left|-1\right\rangle $ components, respectively. The black
dashed (dash-dotted) curve denotes the predictions from Eq.~\eqref{eq:RamDress0}
magnified by $9.1$ in the radial position for $\left|0\right\rangle $
($\left|\pm1\right\rangle $), while the colored dashed curves are
the full numerical simulations of TOF from 3D time-dependent Sch\"{o}rdinger
equation. (b) Radial cross section of the total optical density. Colored
solid and dotted (black dashed) curves indicate the experimental data
(TOF simulation at $\delta=0$). (c) Interference pattern between
$\left|\pm1\right\rangle $ components at $\delta=0$. Adapted from
Ref.~\cite{Chen2018S}.}

\label{fig:LinExp2}
\end{figure}

The vortex structures in the bare hyperfine states could be probed
according to a Stern-Gerlach scheme after a time-of-flight (TOF) expansion
by simultaneously turning off the Raman beams and external trap. The
density profile of bare hyperfine components $\left|m_{F}\right\rangle $
after $24$ms TOF with holding time $1$ms are presented in Fig.~\ref{fig:LinExp2}(a)
for different Raman detunings, as well as the corresponding total
density profile (or optical density) presented in Fig.~\ref{fig:LinExp2}(b).
It is easily found that the $\left|m_{F}=0\right\rangle $ component
carries zero angular momentum, while the $\left|m_{F}=\pm1\right\rangle $
components carry the same magnitude of angular momentum indicated
from their same hole sizes. The interference pattern between $\left|m_{F}=\pm1\right\rangle $
components shown in Fig.~\ref{fig:LinExp2}(c) implies that they carry
the opposite rotation directions of vortices with angular momentum
$l=\mp1$. 

\begin{figure}
\includegraphics[width=1\columnwidth]{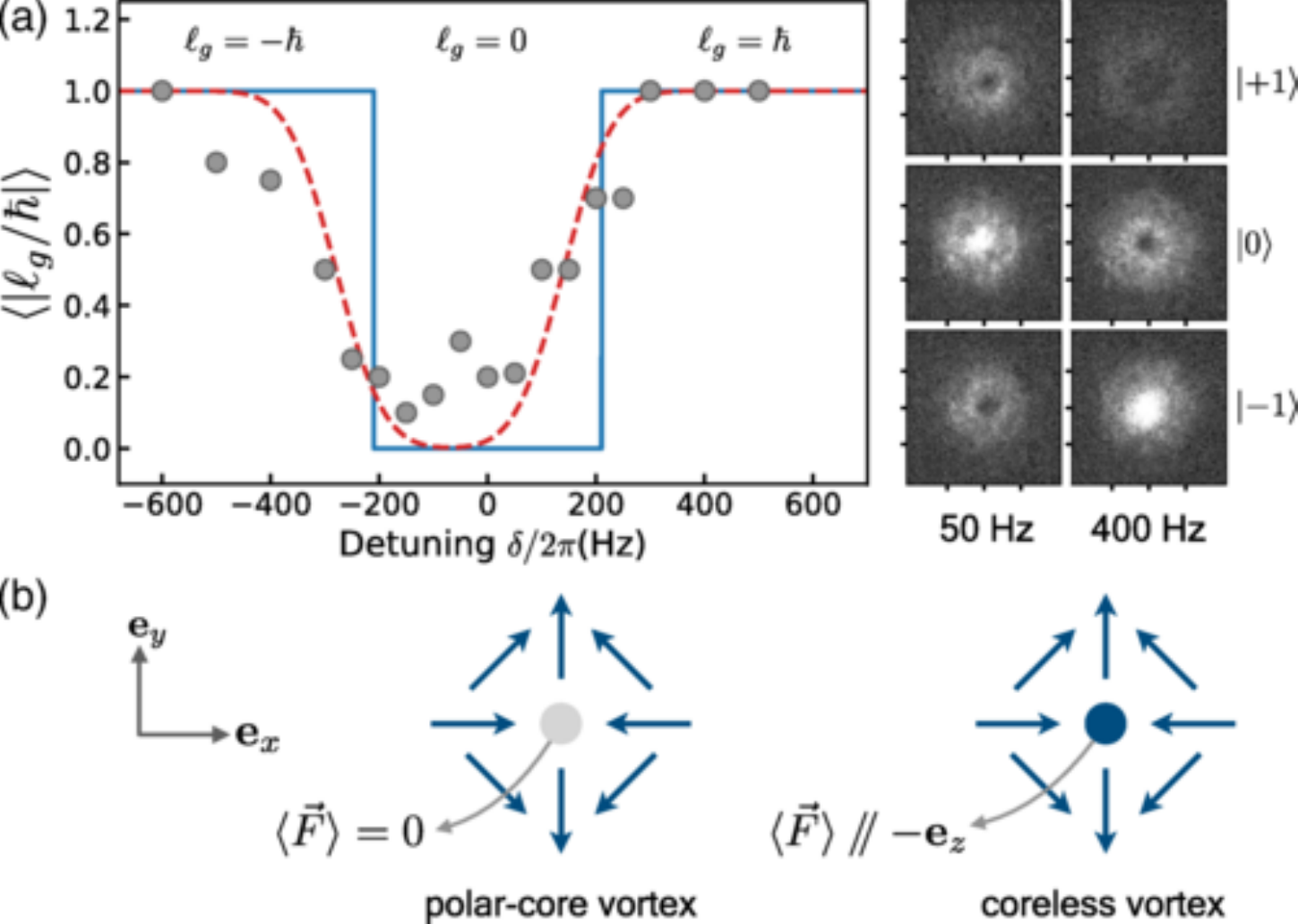}

\caption{(a) Observation of the phase transition of the SOAM-coupled ground
state between $\left|l_{g}\right|=1$ and $l_{g}=0$ states. The solid
and dashed lines are the calculations for the ideal case and that
including the detuning noise in the experiment. Then density profiles
of bare hyperfine states at $\delta/2\pi=50$ and $400$Hz are displayed,
corresponding to the QAMs $l_{g}=0$ and $1$, respectively. (b) Schematic
of the spin textures at $\delta>0$, where the arrows show the direction
of the transverse spin $\left(\left\langle F_{x}\right\rangle ,\left\langle F_{y}\right\rangle \right)$.
Adapted from Ref.~\cite{Chen2018R}.}

\label{fig:LinExp3}
\end{figure}

While the SOAM coupling is demonstrated according to adiabatically
loading atoms into the middle-energy Raman-dressed state $\left|\xi_{0}\right\rangle $,
the effective light-induced gauge potential in this dressed state
is zero, i.e., ${\bf A}_{0}=0$~\cite{Chen2018S}, which leads to
a vanishing magnetic field experienced by atoms, i.e., ${\bf B}=\nabla\times{\bf A}_{0}$.
Subsequently, Lin's group further reports the realization of non-zero
gauge potential with SOAM coupling by adiabatically loading atoms
into the Raman-dressed state $\left|\xi_{-1}\right\rangle $, the
lowest-energy eigenstate of $\boldsymbol{\Omega}_{eff}\cdot{\bf F}$
\citep{Ho1998S,Chen2018R}, 
\begin{multline}
\left|\xi_{-1}\right\rangle =e^{i\left(\theta+\gamma\right)}\left[\frac{e^{i\varphi}}{2}\left(1-\cos\beta\right)\left|-1\right\rangle -\frac{\sin\beta}{\sqrt{2}}\left|0\right\rangle \right.\\
\left.+\frac{e^{-i\phi}}{2}\left(1+\cos\beta\right)\left|+1\right\rangle \right],
\end{multline}
where $\theta+\gamma$ is the phase introduced by a gauge transformation.
Then an azimuthal gauge potential is induced by SOAM coupling, taking
the form of $A_{-1}=\left(\hbar/r\right)\cos\beta$ under the choice
of $\theta+\gamma=0$, which gives rise to an effective magnetic field
experienced by atoms. The magnetic flux through the atomic cloud can
be tuned via the Raman detuning $\delta$. This leads to a phase transition
of the SOAM-coupled ground state from one to another QAM. In the experiment
\citep{Chen2018R}, it is demonstrated that the QAM of the ground
state changes from $l_{g}=\pm1$ to $0$ as the Raman detuning $\left|\delta\right|$
continuously decreases. The phase transition occurs at the critical
Raman detuning $\left|\delta\right|\approx2\pi\times210$Hz as shown
in Fig.~\ref{fig:LinExp3}(a). At a detuning below the critical value,
i.e., $\delta=2\pi\times50$Hz, the system stays at the ground state
with QAM $l_{g}=0$. The corresponding mechanical angular momenta
of bare hyperfine states $\left|m_{F}\right\rangle $ are $l_{m_{F}}=\left(1,0,-1\right)$
for $m_{F}=\left(+1,0,-1\right)$ in the laboratory frame. The same
hole sizes in the $\left|m_{F}=+1\right\rangle $ and $\left|m_{F}=-1\right\rangle $
are observed in the experiment as shown in Fig.~\ref{fig:LinExp3}(a).
At a detuning above the critical value, i.e., $\delta=2\pi\times400$Hz,
the SOAM-coupled ground state transits to the state with QAM $l_{g}=1$,
which corresponds to the mechanical angular momenta of bare hyperfine
states $l_{m_{F}}=\left(2,1,0\right)$ for $m_{F}=+1,0,-1$ in the
laboratory frame. Then the vortex with larger hole size in $\left|m_{F}=+1\right\rangle $,
compared to that in $\left|m_{F}=-1\right\rangle $, is anticipated
and is experimentally confirmed. 

\begin{figure}
\includegraphics[width=1\columnwidth]{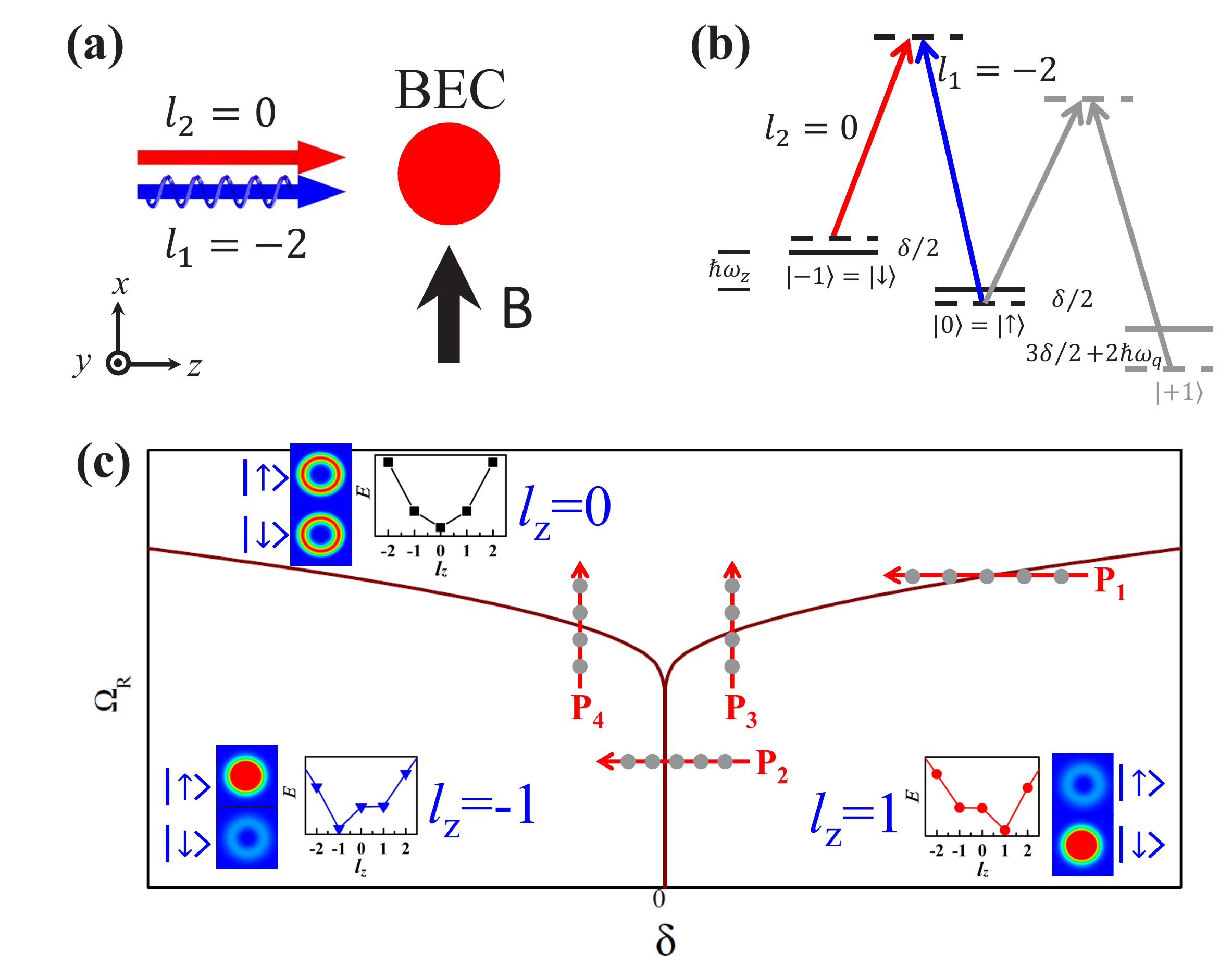}

\caption{Schematic of SOAM coupling. (a) Experimental setup. (b) The energy
diagram of Raman transition. The hyperfine states $\left|m_{F}=0\right\rangle =\left|\uparrow\right\rangle $
and $\left|m_{F}=-1\right\rangle =\left|\downarrow\right\rangle $
are coupled by a pair of Raman beams, one of which is operated in
the LG mode with a winding number $l_{1}=-2$. The third hyperfine
state $\left|m_{F}=+1\right\rangle $ is far-off-resonant due to a
large quadratic Zeeman shift $\omega_{q}\approx2\pi\times5.52$kHz.
Here, $\delta$ is the two-photon detuning. (c) The ground-state phase
diagram. Three single-particle ground-state phases are characterized
by the QAM $l_{z}=0,\pm1$. The corresponding lowest-band dispersion
as well as the typical density profiles is illustrated in the insets.
Here, the phase transitions are studied in the experiment following
four typical paths $P_{1,2,3,4}$. Adapted From Ref.~\cite{Zhang2019G}.}

\label{fig:JiangExp1}
\end{figure}

Soon, the ground-state quantum phase diagram of this SOAM-coupled system is comprehensively studied in an effective spin-half
$^{87}$Rb atomic gas~\cite{Zhang2019G}. Due to a larger quadratic Zeeman shift
$\omega_{q}=2\pi\times5.52$kHz compared to that of~\cite{Chen2018S,Chen2018R},
the third Zeeman hyperfine state $\left|F=1,m_{F}=+1\right\rangle $
is far-off-resonant, and is then dropped out from the problem. In
addition, one of the two Raman beams is operated in a higher LG mode
with a winding number $l_{1}=-2$. It gives rise to a rich ground-state
phase diagram as presented in Fig.~\ref{fig:JiangExp1}. The whole
system is therefore governed by the Hamiltonian \eqref{eq:1Ham1}
in the presence of LG Raman beams, which introduces an effective SOAM
coupling $\hat{L}_{z}\hat{\sigma}_{z}$ as seen in Eq.~\eqref{eq:1Ham2}
after a unitary transformation. Here, a tune-out wavelength of LG
beams is chosen~\cite{Schmidt2016W}. It ensures that any observed
circular structure of atomic clouds is resulted from the vortex formation
due to the SOAM coupling, excluding the trapping effect of the diagonal
ac-stark potential $\chi\left(r\right)$ of LG beams.

In order to explore the ground-state quantum phases of the system,
the Raman beams need to be introduced adiabatically. After preparing
the BEC in the hyperfine state $\left|m_{F}=-1\right\rangle $ by
setting a large detuning $\delta=2\pi\times400$kHz, the Raman beams
are turned on. Then the detuning $\delta$ is ramped to the desired
value adiabatically. This process keeps the system remaining in the
ground state all the time. Subsequently, the transitions between different
ground-state quantum phases could be studied following different paths
$P_{1,2,3,4}$ as shown in Fig.~\ref{fig:JiangExp1}(c). For example,
along the path $P_{1}$, the Raman coupling strength is fixed at $\Omega_{R}/\hbar\omega=1604.5$
in the unit of harmonic energy, while the detuning $\delta$ adiabatically
decreases across the phase boundary. The SOAM-coupled system transits
from the half-skyrmion phase with QAM $l_{z}=1$ to the vortex-antivortex
phase with QAM $l_{z}=0$. The phase transition can experimentally
be identified according to the change of density profiles of bare
hyperfine states: the vortex structure exists only in the spin-$\uparrow$
atomic cloud in the half-skyrmion phase, while both spin states exhibit
vortex formations in the vortex-antivortex phase. The analogous procedures
are performed along the paths $P_{2,3,4}$. 

\begin{figure}
\includegraphics[width=1\columnwidth]{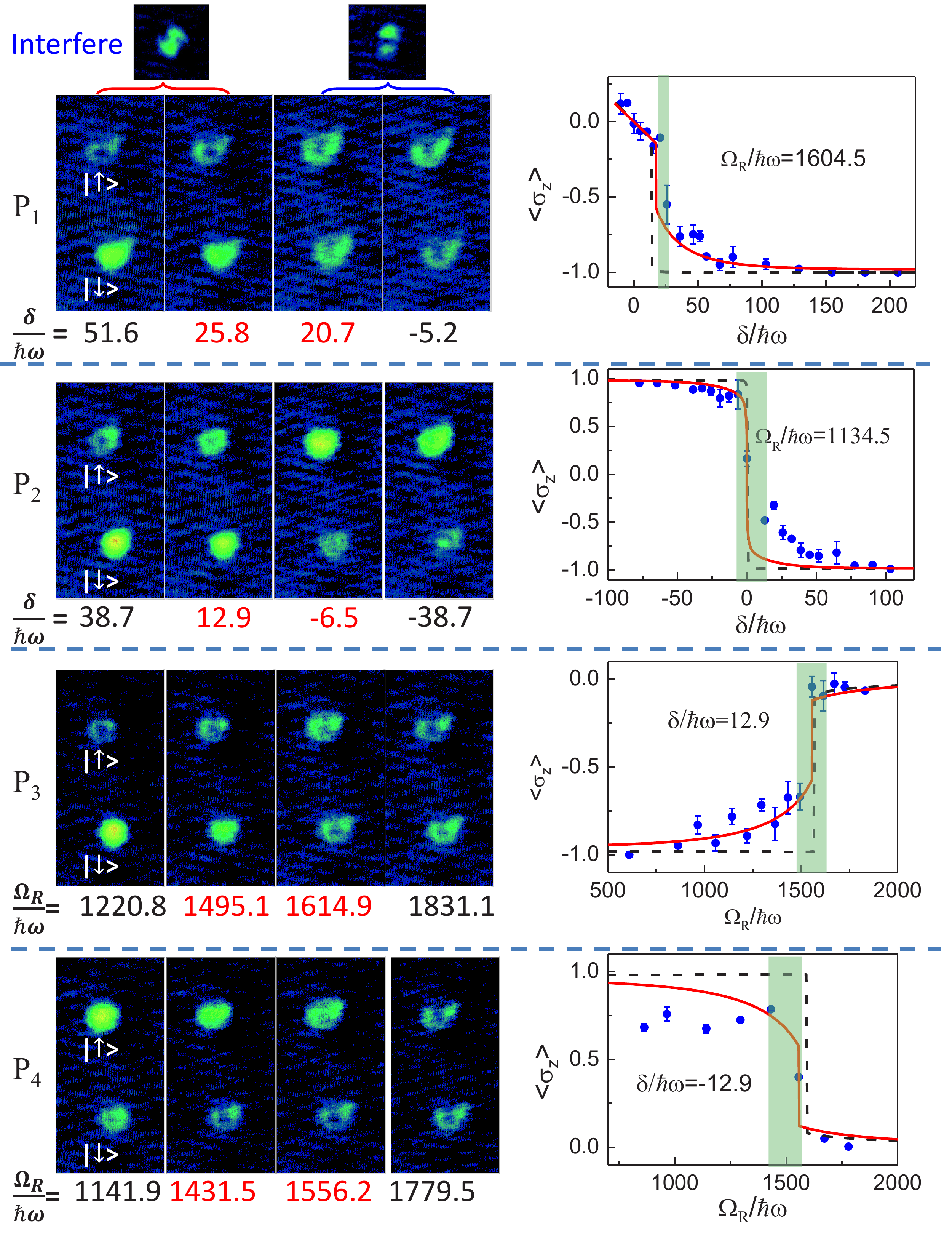}

\caption{The spin-resolved density profiles after a $20$ms TOF expansion across
different phase-transition boundaries along four exemplary paths $P_{1,2,3,4}$
(left panel), and the corresponding spin polarization $\left\langle \hat{\sigma}_{z}\right\rangle $
(right panel). The black dashed (red solid) curve is the theoretical
prediction at zero temperature (at finite temperature $T/T_{c}=0.32$).
Adapted from Ref.~\cite{Zhang2019G}.}

\label{fig:JiangExp2}
\end{figure}

\begin{figure}
\includegraphics[width=1\columnwidth]{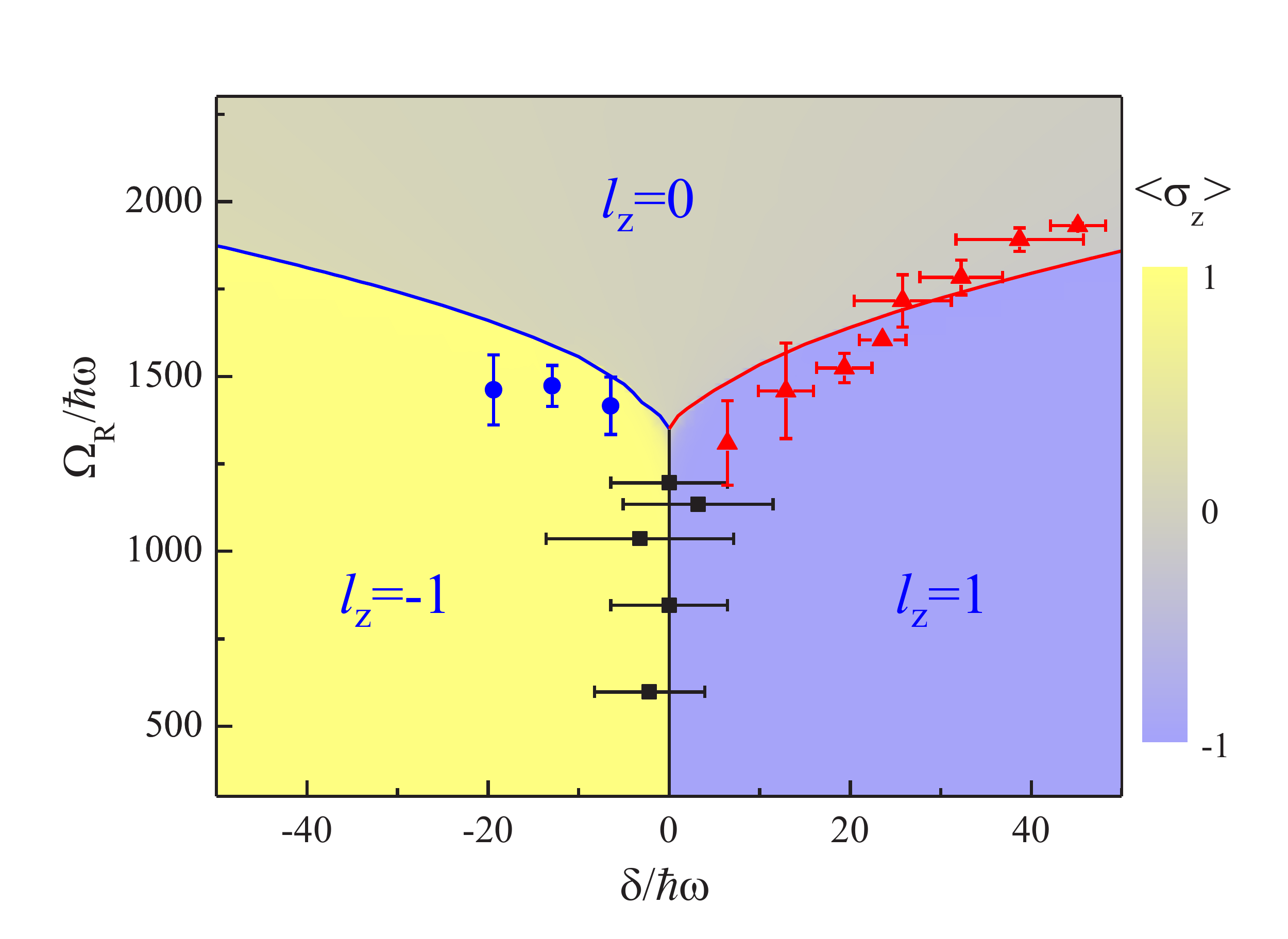}

\caption{The ground-state phase diagram of a SOAM-coupled condensate. Three
quantum phases are characterized by QAMs $l_{z}=0,\pm1$. The solid
curves denote the phase boundaries predicted by the mean-field theory,
which are experimentally confirmed. The background color denotes the
spin polarization $\left\langle \hat{\sigma}_{z}\right\rangle $.
Adapted from Ref.~\cite{Zhang2019G}.}

\label{fig:JiangExp3}
\end{figure}

The spin-resolved density profiles are presented in Fig.~\ref{fig:JiangExp2}
across different phase boundaries. To confirm the formation of vortices
that the circular structures of atom clouds carry angular momenta,
a resonant radio-frequency (rf) pulse is applied, that transfers the
atoms between internal Zeeman hyperfine states $\left|\uparrow\right\rangle $
and $\left|\downarrow\right\rangle $~\cite{Matthews1999V,Andersen2006Q,Wright2009S}.
The appearance of the interference pattern in each spin component
during TOF, as shown in Fig.~\ref{fig:JiangExp2}, implies that the
circular structures are indeed vortices before the rf transition.
The spin polarization $\left\langle \hat{\sigma}_{z}\right\rangle =\left(N_{\uparrow}-N_{\downarrow}\right)/\left(N_{\uparrow}+N_{\downarrow}\right)$
is an additional indicator of the phase transitions, which jumps among
$\left\langle \hat{\sigma}_{z}\right\rangle =0,\pm1$ across different
phase boundaries. Here, $N_{\uparrow\left(\downarrow\right)}$ is
the atom number for the spin-$\uparrow\left(\downarrow\right)$component. The spin polarization is presented as a function of the
two-photon detuning $\delta$ in Fig.~\ref{fig:JiangExp2}. The evident
jump behavior during phase transitions is observed in the experiment
even at finite temperatures. Regarding these characteristic behaviors
across the phase transitions, it allows us to identify the boundaries
of different ground-state quantum phases. Then the ground-state phase
diagram is conveniently mapped out in the parameter space spanned
by the detuning $\delta$ and the coupling strength $\Omega_{R}$,
as illustrated in Fig.~\ref{fig:JiangExp3}.

\section{Conclusions and outlooks}
\label{sec:outlook}
In conclusion, the most recent advances \replaced{in}{on} both the theories and experiments
of spin-orbital-angular-momentum (SOAM) coupled quantum gases are
briefly summarized in this review. The basic idea of engineering SOAM
coupling in ultracold atoms is presented, and exotic quantum phases
of Bose gases are introduced as well as those of Fermi gases in the
presence of SOAM coupling. In spite of remarkable progress in this
field, it is far from the end of the story. There could be further
developments in SOAM-coupled quantum gases, since several issues
still require more theoretical and experimental efforts.

\emph{Few-body physics}---As a building block of interacting many-body
systems, the few-body problem is of important significance in ultracold
atoms. For instance, the two-body physics determines the essential
interaction parameter in the many-body Hamiltonian, and even dominates
crucial quantum correlations in many-body systems, such as Tan's relations
\citep{Tan2008E,Tan2008L,Tan2008G}. Besides, the few-body problem
itself displays intriguing features such as the Feshbach resonance
and Efimov effect. The few-body problem has intensively been studied
in spin-linear-momentum-coupled quantum gases~\cite{Cui2012M,Zhang2012M,Cui2014U,Shi2014U,Peng2018C,Zhang2020U,Zhang2021U},
while the theory for SOAM-coupled systems is still elusive to date.
Rich physics in this new few-body system is remained to be discovered
in near future, which could provide deep insight \replaced{into}{of} many-body properties
of SOAM-coupled quantum gases.

\emph{Finite-temperature phase diagram}s---The current theory of
SOAM-coupled quantum systems is constructed in the framework of the
zero-temperature mean-field theory. For example, the ground-state
quantum phases of Bose gases are based on the solution of \added{the} Gross-Pitaevskii
equation, while the giant vortex Fermi superfluid is predicted by
the zero-temperature Bogoliubov-de Gennes \replaced{formalism}{formulism}. The natural question
arises whether these exotic quantum phases are stable enough against
the thermal fluctuation at finite temperature as well as against the
quantum fluctuation at zero temperature beyond the mean-field theory \cite{Chen2017Q}.
A more comprehensive theoretical analysis is required to address these
issues, and is necessary for the comparison with experimental observations as well.

\emph{Experimental perspectives}---The angular stripe phases of SOAM-coupled
Bose gases have \added{not} yet been observed in experiments. It is energetically
favored for an intraspecies interaction strength larger than the interspecies
one (i.e., $g_{\uparrow\uparrow},g_{\downarrow\downarrow}>g_{\uparrow\downarrow}$).
However, there is no convenient Feshbach resonance of $^{87}$Rb
atoms to adjust interactions. The angular stripe phase lies in a narrow
parameter window due to the small difference between $g_{\uparrow\uparrow}$
(or $g_{\downarrow\downarrow}$) and $g_{\uparrow\downarrow}$ for
$^{87}$Rb atomic gases. For this reason, the $^{41}$K atomic gas
becomes one of \added{the} promising candidates for future experiments~\cite{Chen2020A},
in which the interspecies scattering length can be tuned in a wide
range near the Feshbach resonance centered at the magnetic field $B_{0}=51.95$G,
while the intraspecies scattering lengths are approximately constant.
By choosing appropriately the interatomic scattering lengths, the
window of angular stripe phases in the parameter space is enlarged,
which can be operated easily in experiments~\cite{Chen2020A}.

Moreover, the SOAM coupling is not realized in Fermi gases at present,
which would provide a new platform to investigate exotic vortex and
topological superfluid states. Other interesting directions, such
as the non-equilibrium or the dynamics, BCS-BEC crossover, \added{and} quantum
fluctuations of Fermi gases with SOAM coupling are yet to \replaced{be explored}{explore}.
All these fascinating and remarkable phenomena are devoted to future
studies.

\section{acknowledgments}
We are grateful for inspiring discussions with Tian-You Gao, Wei
Yi and Fan Wu. SGP and KJ are supported by the National Natural Science Foundation
of China under Grant Nos. 11974384 and 12121004, \added{the National Key R\&D Program under
Grant No. 2022YFA1404102}, Chinese Academy of Sciences under Grant
 No. YJKYYQ20170025, K. C. Wong Education Foundation under Grant No. GJTD-2019-15,
and the Natural Science Foundation of Hubei Province under Grant No. 2021CFA027. 
XLC is supported by \added{the Natural Science Foundation of
China under Grant No. 12204413 and} the Science Foundation of Zhejiang Sci-Tech University under Grant No. 21062339-Y. KJC is supported by the Natural Science Foundation of
China under Grant No. 12104406 and the Science Foundation of Zhejiang
Sci-Tech University under Grant No. 21062338-Y. PZ is supported by
the National Natural Science Foundation of China under Grant No. 11804177.
LYH is supported by the National Key R\&D Program under Grant No. 2018YFA0306503.

\end{document}